\documentclass[twocolumn]{IEEEtran}
\usepackage{amsmath,amssymb,amsthm,epsfig,color,subfigure,empheq,graphicx,graphics,balance}
\usepackage{enumerate,url,algorithm,algorithmic,wasysym,epstopdf,enumitem,array}
\usepackage{xcolor}
\usepackage{amsfonts}
\usepackage{adjustbox}
\usepackage{multirow}
\usepackage{accents}
\newcommand{\ubar}[1]{\underaccent{\bar}{#1}}

\DeclareMathOperator{\nullspace}{null}

\DeclareMathOperator{\diag}{dg}

\newtheorem{remark}{Remark}

\newcommand \bzero{\mathbf{0}}
\newcommand \bone{\mathbf{1}}

\newcommand \bb{\mathbf{b}}
\newcommand \bc{\mathbf{c}}

\newcommand \bef{\mathbf{f}} 

\newcommand \bp{\mathbf{p}}
\newcommand \bq{\mathbf{q}}

\newcommand \bs{\mathbf{s}}

\newcommand \bv{\mathbf{v}}

\newcommand \bx{\mathbf{x}}
\newcommand \by{\mathbf{y}}
\newcommand \bz{\mathbf{z}}
\newcommand \bA{\mathbf{A}}
\newcommand \bB{\mathbf{B}}

\newcommand \bD{\mathbf{D}}

\newcommand \bI{\mathbf{I}}

\newcommand \bR{\mathbf{R}}
\newcommand \bS{\mathbf{S}}

\newcommand \bU{\mathbf{U}}
\newcommand \bV{\mathbf{V}}

\newcommand \bX{\mathbf{X}}

\newcommand \bZ{\mathbf{Z}}

\newcommand \bdelta{\boldsymbol{\delta}}

\newcommand \bzeta{\boldsymbol{\zeta}}
\newcommand \btheta{\boldsymbol{\theta}}

\newcommand \blambda{\boldsymbol{\lambda}}
\newcommand \bmu{\boldsymbol{\mu}}
\newcommand \bnu{\boldsymbol{\nu}}

\newcommand \bSigma{\mathbf{\Sigma}}

\newcommand \mcC{\mathcal{C}}

\newcommand \mcL{\mathcal{L}}

\newcommand \mcN{\mathcal{N}}
\newcommand \mcO{\mathcal{O}}

\newcommand \mcS{\mathcal{S}}
\newcommand \mcT{\mathcal{T}}

\newcommand \hbSigma{\hat{\mathbf{\Sigma}}}

\newcommand \bby{\bar{\mathbf{y}}}

\newcommand \bbZ{\bar{\mathbf{Z}}}

\newcommand \bbnu{\bar{\boldsymbol{\nu}}}

\renewcommand{\d}[1]{\ensuremath{\operatorname{d}\!{#1}}}

\begin{document}

\title{Fast Inverter Control by Learning the OPF Mapping using Sensitivity-Informed Gaussian Processes}

\author{
	Mana Jalali,~\IEEEmembership{Graduate Student Member,~IEEE},
    Manish K. Singh,~\IEEEmembership{Member,~IEEE}, 
	Vassilis Kekatos,~\IEEEmembership{Senior Member,~IEEE}, 
	Georgios B. Giannakis,~\IEEEmembership{Fellow,~IEEE}, and 
	Chen-Ching Liu,~\IEEEmembership{Fellow,~IEEE},
	

\thanks{M.~Jalali, V.~Kekatos, and C.-C. Liu are with the Bradley Dept. of ECE, Virginia Tech, Blacksburg, VA 24061, USA. M. K. Singh and G. B. Giannakis are with the Un. of Minnesota, Minneapolis, MN 55455. Emails: manaj2@vt.edu, msingh@umn.edu, kekatos@vt.edu, georgios@umn.edu, and ccliu@vt.edu. This work was supported by the U.S. National Science Foundation grants 1751085 and 2034137, and the Commonwealth Cyber Initiative (CCI) Southwest Node, State of Virginia, USA.}	


\thanks{Digital Object Identifier XXXXXX}
}	
	

\maketitle

\begin{abstract}
Fast inverter control is a desideratum towards the smoother integration of renewables. Adjusting inverter injection setpoints for distributed energy resources can be an effective grid control mechanism. However, finding such setpoints optimally requires solving an optimal power flow (OPF), which can be computationally taxing in real time. Previous works have proposed learning the mapping from grid conditions to OPF minimizers using Gaussian processes (GPs). This GP-OPF model predicts inverter setpoints when presented with a new instance of grid conditions. Training enjoys closed-form expressions, and GP-OPF predictions come with confidence intervals. To improve upon data efficiency, we uniquely incorporate the sensitivities (partial derivatives) of the OPF mapping into GP-OPF. This expedites the process of generating a training dataset as fewer OPF instances need to be solved to attain the same accuracy. To further reduce computational efficiency, we approximate the kernel function of GP-OPF leveraging the concept of random features, which is neatly extended to sensitivity data. We perform sensitivity analysis for the second-order cone program (SOCP) relaxation of the OPF, whose sensitivities can be computed by merely solving a system of linear equations. Extensive numerical tests using real-world data on the IEEE 13- and 123-bus feeders corroborate the merits of GP-OPF.
\end{abstract}
	


\section{Introduction}\label{sec:intro}
\allowdisplaybreaks
Rapid fluctuations in power injections by solar photovoltaics and other distributed energy resources (DERs) induce undesirable voltage deviations in distribution grids. Reactive power compensation and active power curtailment by the smart inverters interfacing DERs have been suggested as an effective fast-responding voltage control mechanism. Nonetheless, finding the optimal power injection setpoints for hundreds of inverters is a computationally formidable task. It requires solving the OPF in near real time to account for varying solar and loading conditions. To expedite optimal inverter control, this work aims at learning the OPF mapping using GPs.

Per the IEEE 1547 standard, inverter setpoints can be selected upon Volt-VAR, Watt-VAR, or Volt-Watt curves driven by local data~\cite{Turitsyn11}. Although such rules have been shown to be stable, their equilibria may not be optimal~\cite{9091863}, \cite{VKZG16}, or even perform worse than the no-reactive support option~\cite{Jabr18}. Given the current grid conditions, inverter setpoints can be optimally decided upon solving an OPF. Albeit non-convex, the OPF can be relaxed to a convex second-order cone program (SOCP) or a semidefinite program (SDP); see \cite{Low14} for a survey. To reduce computational time, one may resort to a linearized feeder model and trade modeling accuracy for complexity, to eventually express the OPF as a linear or quadratic program (LP/QP); see e.g., \cite{TJKT20} and references therein. 

To further expedite optimal inverter control, recent works rely on machine learning (ML) techniques to shift some of the computational burden of the OPF from real time to offline. According to this \emph{learning-to-optimize} paradigm, kernel--based regression has been utilized to learn inverter control rules in~\cite{Kara18} and \cite{jalaliSVM}. Deep neural networks (DNNs) have been trained to predict OPF solutions under linearized~\cite{DeepOPFPan19}, \cite{pascal21}, and the exact AC grid models~\cite{ZamzamBaker19}, \cite{GuhaACOPF}. The utility may be interested in the average or probabilistic performance of inverters across a range of uncertain grid conditions. Under this stochastic setup, inverter control rules have been obtained using reinforcement learning strategies~\cite{Nanpeng19}, \cite{ZhangISU}. References \cite{OPFandLearnTSG21} and \cite{GKJ2020} develop inverter control rules driven by incomplete or noisy data, by training a DNN using primal/dual updates based on the Lagrangian function of a stochastic OPF.


Under the deterministic setup, learning-to-optimize schemes proceed in two steps: \emph{s1)} They first solve a large number of OPFs to build a labeled dataset; and \emph{s2)} Train the ML model. Although it occurs offline, training has to be repeated afresh with any alteration of the OPF structure due to a topology reconfiguration or when a large DER goes offline. To expedite \emph{s1)}, in \cite{SGKCB2020} and \cite{L2O2021}, we proposed training a DNN to match not only the OPF solutions, but also their partial derivatives with respect to grid conditions. Owing to this sensitivity-informed training, the DNN attained the same prediction accuracy using a much smaller training dataset. Therefore step \emph{s1)} required solving fewer OPF instances. Reference~\cite{Fioretto1} trained DNNs to learn OPF minimizers by properly penalizing optimality conditions. The latter approach was combined with sensitivity-informed learning in~\cite{Spyros21}. Reference~\cite{SELM20} leverages multi-parametric programming to classify OPF inputs and design DNNs with reduced complexity for each class. In any case, training a DNN during \emph{s2)} itself requires solving an optimization. Moreover, DNN-based OPF predictions come without confidence intervals. To overcome these shortcomings, we rely on modeling the OPF mapping using GPs previously suggested in \cite{GP4UQ} for dealing with a probabilistic OPF.

{The motivation for using GPs is twofold: First, the weights of a GP model can be computed in closed-form, while the few kernel parameters involved are found by solving small-scale minimizations. Second, by virtue of its Bayesian nature, a GP model not only predicts an OPF minimizer, but can also quantify the uncertainty of such prediction in the form of variance. This is important as it may warn the operator not to trust a specific prediction and opt for solving one more instance of the OPF instead. Uncertainty quantification can also be used to identify undersampled areas of the parameter space of the OPF, or areas where the OPF mapping is more complex. Then, additional samples from those areas can be drawn to be solved to strategically enrich the dataset.}

{GPs have been utilized in the power systems literature before. For example, \cite{GP4dynamicsPRWRS21} and \cite{GP4dynamicsCDC21} infer frequency oscillations from synchrophasor data in transmission systems using GPs. For distribution grids, the inverse power flow mapping from power injections to voltages in active distribution grids is modeled as a GP in~\cite{GP4PF}. Reference~\cite{LinearGP4VoltageControl} pursues inverter-based voltage control by approximating voltages as affine functions of injections using a GP having a linear kernel. GP learning has also been used previously for uncertainty propagation through the OPF in~\cite{GP4UQ}. Building on \cite{GP4UQ}, we suggest learning the deterministic OPF mapping in a physics-informed manner by uniquely incorporating the sensitivities of the OPF solutions with respect to problem parameters, i.e., the grid conditions.} 

While the proposed use of OPF sensitivities for improving GP-OPF estimates is novel, there has been significant interest in computing such sensitivities for other applications~\cite{Almeida94parametric}, \cite{Ajjarapu95OCPF}. These works exploited OPF sensitivities to efficiently compute OPF minimizers and look into binding constraints for a given trajectory of load variations. Thus, these works confined their focus on scalar parameterization of loads. Beyond the power systems literature, extensive developments have been reported in the general area of sensitivity analysis of continuous optimization problems~\cite{shapiro2013perturbation}, \cite{Conejo06}. Building upon these seminal works, and relaxing some of their underlying assumptions, a convenient approach for sensitivity analysis of the OPF has been recently proposed in~\cite{L2O2021}. However, the developed approach applies to continuous optimization problems with twice-differentiable \emph{scalar} constraint functions. Although differentiating through convex cone constraints is possible~\cite{agrawal2020differentiating}, the approach gets more involved. Fortunately, simple reformulations allow us for the first time to compute the sensitivities for the minimizers of the SOCP-based OPF. 

{Adopting the idea of \cite{GP4UQ} from the probabilistic to the deterministic setup, this work learns the deterministic OPF mapping using GPs for near-optimal real-time inverter control. As with \cite{GP4UQ}, during training, the GP-OPF model is computed in closed-form. During operation, GP-OPF provides point predictions and confidence intervals for OPF solutions also in closed-form (Section~\ref{sec:GP-OPF}). Beyond the application setup, the technical contribution of this work is on three fronts:}

\emph{i)} We incorporate the sensitivities of OPF solutions with respect to OPF parameters to arrive at a \emph{sensitivity-informed} GP-OPF (SI-GP-OPF). It essentially augments labeled data per solved OPF instance, and can thus attain the same prediction accuracy given a smaller dataset. It thus reduces the number of OPFs to be solved during training (Section~\ref{sec:SI-GP-OPF}). 

\emph{ii)} To reduce computational complexity, we approximate kernel functions using the concept of \emph{random features (RF)} and obtain an RF-based GP-OPF (RF-GP-OPF). Random features are neatly extended to OPF sensitivities to derive an RF-SI-GP-OPF (Section~\ref{sec:rf}). 

\emph{iii)} We perform sensitivity analysis of the SOCP-OPF with respect to load demand and solar generation. Finding the sensitivities of OPF solutions is as easy as solving a system of linear equations (Section~\ref{sec:sa}). 

Extensive numerical tests on the IEEE 13- and 123-bus feeders corroborate our findings (Section~\ref{sec:tests}).


{\emph{Notation:} Column vectors (matrices) are denoted by lower- (upper-) case letters. Symbol $(\cdot)^\top$ stands for transposition; $\bI_{N}$ is the $N \times N$ identity matrix; $\mathbb{E}$ is the expectation operator; and $\|\bx\|$ is the $\ell_2$-norm of vector $\bx$.}

\section{Optimal Inverter Control}\label{sec:SI-OPF}
Consider a distribution feeder with $N+1$ buses hosting a combination of inelastic loads and DERs. Buses are indexed by set $\mcN:=\{1,\dots,N\}$, while the substation is indexed as $n=0$. Suppose there are $N_g$ buses hosting DERs and their indexes are collected in set $\mcN_g \subseteq \mcN$. Let $v_n$ denote the squared voltage magnitude at bus $n$, and $p_n + j q_n$ the complex power injected at bus $n$. Power injections at buses in $\mcN_g$ can be decomposed to controllable inverter-interfaced DER generation and uncontrollable loads as
\begin{equation}\label{eq:pq}
    p_n = p_n^g - p_n^\ell\quad \text{and}\quad q_n = q_n^g - q_n^\ell, \quad \forall n \in \mcN_g.
\end{equation}

Given load demands $\{(p_n^\ell,q_n^\ell)\}_{n\in\mcN}$ at all buses, the task of optimal inverter control amounts to finding the setpoints $\{(p_n^g,q_n^g)\}_{n\in\mcN_g}$ for inverter power injections to minimize a desirable objective while adhering to feeder and inverter ratings. Before particularizing the related optimization problem, we briefly review the \emph{DistFlow} model~\cite{TJKT20}. This model applies to radial single-phase feeders, but can approximate radial three-phase primary networks under nearly balanced operation. 

On a radial single-phase feeder, each bus $n$ has a unique parent $\pi_n$ and a set of children buses $\mcC_n$. The line connecting bus $n$ to its parent bus $\pi_n$ is indexed by $n$. For line $n$, the impedance is denoted by $r_n+jx_n$; the squared magnitude of its current by $\ell_n$; and the complex power flow seen at $\pi_n$ by $P_n+jQ_n$. The feeder topology is assumed known and remains fixed. According to \emph{DistFlow}, the feeder is governed by the ensuing equations for all $n \in \mcN$~\cite{TJKT20}:
\begin{subequations}\label{eq:DF}
\begin{align}\label{eq:DF:p}
    p_n &= \sum_{k \in \mcC_n} P_k - (P_n - r_n \ell_n)\\\label{eq:DF:q}
    q_n &= \sum_{k \in \mcC_n} Q_k - (Q_n - x_n \ell_n)\\\label{eq:DF:v}
    v_n &= v_{\pi_n} + (r_n^2 + x_n^2)\ell_n  - 2\left(r_n P_n + x_n Q_n\right)\\\label{eq:DF:ell}
    \ell_n &= \frac{P_n^2 + Q_n^2}{v_{\pi_n}}.
\end{align}
\end{subequations}


Given the aforesaid model, we next pose a possible rendition of the optimal inverter control task. A utility could determine inverter setpoints by solving the OPF problem:
\begin{subequations}\label{eq:OPF}
\begin{align}
    \min ~~& \sum_{n=1}^N \left(p_n + r_n \ell_n\right)\\
    \textrm{over}~~&\bx:=\{\{p_n^g, q_n^g\}_{n \in \mcN_g}, \{P_n, Q_n, v_n, \ell_n\}_{n=1}^N, v_0\}\label{eq:OPF:x}\\
    \textrm{s.to}~~& \eqref{eq:pq},\eqref{eq:DF:p}-\eqref{eq:DF:v}\label{eq:OPF:eq}\\
    &\left\|\begin{bmatrix}
    2P_n\\
    2Q_n\\
    v_{\pi_n} - \ell_n
    \end{bmatrix}\right\| \leq \ell_n + v_{\pi_n}, \quad n \in \mcN\label{eq:OPF:soc}\\
    ~&~\ell_n \leq \bar{\ell}_n, \quad n \in \mcN\label{eq:OPF:lmax}\\
    ~&~\ubar{v}_n \leq v_n \leq \bar{v}_n, \quad n \in \mcN\label{eq:OPF:vmax}\\
    ~&~0 \leq  p_n^g \leq \bar{p}_n^g, \quad n \in \mcN_g\label{eq:OPF:pmax}\\
    ~&~\left(p_n^g\right)^2 + \left( q_n^g\right)^2 \leq \left( \bar{s}_n^g \right)^2, \quad n \in \mcN_g\label{eq:OPF:smax}
\end{align}
\end{subequations}
{over optimization variable $\bx$.} Problem~\eqref{eq:OPF} aims at minimizing the total active power flowing via the substation to the feeder including the ohmic losses along all lines. The second-order cone (SOC) constraint in \eqref{eq:OPF:soc} constitutes the widely adopted convex relaxation of \eqref{eq:DF:ell}~\cite{FL1}. For several renditions of the OPF, this relaxation has been shown to be \emph{exact}~\cite{Low14}, that is \eqref{eq:OPF:soc} holds with equality at optimality for all $n$. Constraint \eqref{eq:OPF:lmax} ensures currents remain below the prescribed line ampacities $\bar{\ell}_n$.  Constraint \eqref{eq:OPF:vmax} maintains voltages within a given range $[\ubar{v}_n,\bar{v}_n]$ for all $n\in\mcN$. Constraint \eqref{eq:OPF:pmax} limits the active power generated by inverter $n$ to remain smaller than or equal to the maximum possible value $\bar{p}_n^g$, which varies across time as it depends on the currently experienced solar irradiance. Finally, constraint \eqref{eq:OPF:smax} limits the apparent power of inverter $n$ depending on its kVA rating $\bar{s}_n^g$. Although this is a convex quadratic constraint, we cast it in the SOC form $\|\left[p_n^g~~q_n^g\right]^\top\|\leq \bar{s}_n^g$ to unify the exposition. To keep the presentation uncluttered, it is assumed that each bus hosts at most one inverter. 

The OPF in \eqref{eq:OPF} constitutes a \emph{parametric} optimization problem, which has to be solved every time load demands and solar generation change. Collect the OPF parameters in vector
\begin{equation}\label{eq:btheta}
\btheta:=\left(\{p_n^\ell, q_n^\ell\}_{n=1}^N,\{\bar{p}_n^g\}_{n \in \mcN_g}\right).
\end{equation}
The length of this parameter vector is $M:=2N+N_g$. We can now pose \eqref{eq:OPF} as the parametric SOCP
\begin{subequations}\label{eq:OPF2}
\begin{align}
    \bx_{\btheta}:=\arg\min_{\bx}~~&~\bc^\top \bx\\ 
    \textrm{s.to}~~&~\bA_e \bx  = \bB_e \btheta + \bef_e\label{eq:OPF2:eq}\\
    ~&~\bA_i \bx \leq \bB_i \btheta + \bef_i\label{eq:OPF2:ineq}\\
    ~&~\|\bA_m \bx\| \leq \bb_m^\top \bx  + f_m, ~ m=1:2N.\label{eq:OPF2:soc}
\end{align}
\end{subequations}
Constraint~\eqref{eq:OPF2:eq} collects the equality constraints in \eqref{eq:OPF:eq}. Constraint~\eqref{eq:OPF2:ineq} collects the inequality constraints in \eqref{eq:OPF:lmax}--\eqref{eq:OPF:pmax}. Constraint~\eqref{eq:OPF2:soc} collects constraints \eqref{eq:OPF:soc}, \eqref{eq:OPF:smax}. The involved matrices $(\bA_e,\bA_i,\bA_m,\bB_e,\bB_i)$, vectors $(\bc,\bef_e,\bef_i,\bb_m)$, and scalars $f_m$ follow directly from \eqref{eq:OPF}. Let $\bx_{\btheta}$ denote the minimizer of the OPF associated with parameter $\btheta$.

The goal of this work is to learn the \emph{mapping} $\btheta\rightarrow\bx_{\btheta}$ induced by the SOCP-based OPF in \eqref{eq:OPF2}. We would like to train a machine learning model that once presented a $\btheta$, it predicts the associated minimizer $\bx_{\btheta}$. Such model is useful in different applications. For example, predictions of $\bx_{\btheta}$ can be directly used as setpoints, thus accelerating the task of inverter control. Alternatively, they can be used to warm-start an OPF solver. OPF predictions can also be used in hosting capacity analyses where a system operator studies different levels of renewable integration and the approximate effect of inverter control. Depending on the application, quantifying the uncertainty for any given OPF prediction can be also important. Training such model involves three phases: \emph{i)} Creating a labeled dataset by solving \eqref{eq:OPF2} for different $\btheta$'s to find the related $\bx_{\btheta}$'s; \emph{ii)} Training a learning model using the labeled dataset offline; and \emph{iii)} Using the trained model in real-time to predict OPF solutions.

\section{Modeling the OPF Mapping as a GP}\label{sec:GP-OPF}
{Reference~\cite{GP4UQ} suggested modeling mapping $\btheta\rightarrow\bx_{\btheta}$ as a GP to deal with a probabilistic OPF, i.e., to efficiently approximate empirical histograms of OPF solutions over randomly sampled grid conditions $\btheta$'s. Spurred by \cite{GP4UQ}, we propose a GP-OPF model for expediting real-time near-optimal inverter control. This section reviews GP-OPF from \cite{GP4UQ} under the inverter control setup. Over the following sections we improve upon \cite{GP4UQ} towards: \emph{1)} including OPF sensitivities to enhance data efficiency; \emph{2)} implementing GP using random features to improve on computational complexity during training; and \emph{3)} accomplishing sensitivity analysis of the SOCP-based OPF.}

Towards learning the OPF mapping, the entries of $\bx_{\btheta}$ are learned independently. We henceforth focus on a particular entry, say the injection $q_n^g$ by inverter $n$. To simplify notation, we denote this entry as $y(\btheta)$. By sampling $T$ loading conditions $\{\btheta_t\}_{t=1}^T$, we solve \eqref{eq:OPF2} to optimality. We have thus constructed a labeled training dataset $\mcT:=\{\left(\btheta_t,y(\btheta_t)\right)\}_{t=1}^T$. Using this dataset, our goal is to learn the function $y:\mathbb{R}^M\rightarrow \mathbb{R}$ so we are able to predict $y(\btheta)$ for unseen values of $\btheta$. The function $y(\btheta)$ will be learned using GP regression, which is GP-OPF as briefly review next.

GP-OPF relies on a key property of the multivariate Gaussian probability density function (PDF). Consider a random vector $\by\sim \mcN(\bmu,\bSigma)$ drawn from a Gaussian PDF with mean $\bmu$ and covariance $\bSigma$. Partition vector $\by$ into two subvectors as
\begin{equation}\label{eq:GP}
    \by=\begin{bmatrix}
    \by_1\\
    \by_2
    \end{bmatrix} \sim \mcN \left( \begin{bmatrix}
    \bmu_1\\
    \bmu_2
    \end{bmatrix}, \begin{bmatrix}
    \bSigma_{11} & \bSigma_{21}^\top\\
    \bSigma_{21} & \bSigma_{22}
    \end{bmatrix} \right)
\end{equation}
where $\bmu$ and $\bSigma$ have been partitioned conformably. Because subvectors $\by_1$ and $\by_2$ are jointly Gaussian, the conditional PDF of $\by_2$ given $\by_1$ is also Gaussian with mean and covariance
\begin{subequations}\label{eq:mmse}
\begin{align}\label{eq:mmse:m}
    \mathbb{E}[\by_2|\by_1] &= \bmu_2 + \bSigma_{21} \bSigma_{11}^{-1}\left(\by_1-\bmu_1\right)\\ \label{eq:mmse:c}
    \mathrm{Cov}[\by_2|\by_1]  &= \bSigma_{22} -  \bSigma_{21} \bSigma_{11}^{-1} \bSigma_{21}^\top.
\end{align}
\end{subequations}
The implication is that if $\by_1$ is known and $\by_2$ is not, then $\mathbb{E}[\by_2|\by_1]$ provides an estimate for $\by_2$. This is in fact the minimum mean squared error (MMSE) estimate of $\by_2$. In addition to the point estimate of \eqref{eq:mmse:m}, the covariance $\mathrm{Cov}[\by_2|\by_1]$ quantifies the uncertainty of this estimate, which can be used to provide confidence intervals~{~\cite[Ch.~2]{GP},~\cite[Ch.~5]{Bishop}}. 

{The OPF mapping can be learned by modeling $y(\btheta)$ as a GP over $\btheta$ as suggested in~\cite{GP4UQ}.} A random process is a GP if a collection of any number of samples forms a Gaussian random vector~\cite[Ch.~1]{GP}. In other words, function $y(\btheta)$ is a GP if any vector $\by$ having entries $y(\btheta_i)$ over any collection of $\btheta_i$'s follows a Gaussian PDF as in \eqref{eq:GP}. The idea is to let subvector $\by_1$ collect the already computed OPF solutions $\{y(\btheta_t)\}_{t=1}^T$ from dataset $\mcT$, and subvector $\by_2$ collect the solutions we would like to infer and correspond to parameter vectors $\mcS:=\{\btheta_s\}_{s=1}^S$. Thanks to \eqref{eq:mmse}, we can use the training dataset $\mcT$ to predict OPF decisions over any testing dataset $\mcS$.

For \eqref{eq:GP}--\eqref{eq:mmse} to be useful for any reasonable $\btheta$'s in $\mcT$ and $\mcS$, GP regression parameterizes the mean and covariance of $y(\btheta)$ as a function of $\btheta$. The mean is typically modeled as zero, that is $\mu(\btheta_i)=0$ for all $\btheta_i$. This is without loss of generality as explained in~\cite{GP}. As for the covariance matrix $\bSigma$, note that its $(i,j)$-th entry corresponds to the covariance $\mathbb{E}[y(\btheta_i)y(\btheta_j)]$. In GP-OPF, the latter is assumed to be described as $\mathbb{E}[y(\btheta_i)y(\btheta_j)]=k(\btheta_i,\btheta_j)$, where $k(\btheta_i,\btheta_j)$ is a \emph{kernel function} measuring the similarity between any two parameter vectors. A common choice is the Gaussian kernel
\begin{equation}\label{eq:kernel}
k(\btheta_i,\btheta_j)= \alpha e^{-\frac{\beta}{2}\|\btheta_i-\btheta_j\|^2}
\end{equation}
for positive $\alpha$ and $\beta$. The Gaussian kernel is a \emph{shift-invariant} kernel as it measures the similarity between two vectors as a function of their distance alone as $k(\btheta_i,\btheta_j)=k(\btheta_i-\btheta_j)$. 

Although the kernel captures the covariance of the actual function, vector $\by_1$ entails observing functions under noise. In our learning-to-optimize setup, the OPF labels are not corrupted by measurement noise (unless the solver was terminated prematurely). However, they do come with modeling noise as the postulated GP model may not be able to match the OPF mapping perfectly. Modeling noise is assumed to be drawn independently from a zero-mean Gaussian PDF with variance $\gamma>0$. Then, the $(i,j)$-th entry of $\bSigma_{11}$ in \eqref{eq:mmse} is given by
\begin{equation}\label{eq:kernel+noise}
\left[\bSigma_{11}\right]_{i,j}=k(\btheta_i,\btheta_j)+\gamma\delta_{ij}
\end{equation}
where $\delta_{ij}$ is the Kronecker delta function. Parameters $(\alpha,\beta,\gamma)$ can be found via maximum likelihood estimation (MLE) using dataset $\mcT$; see~\cite[Ch.~5]{GP}. This is possible since $\bSigma_{11}$ and $\by_1\sim \mcN(\bmu_1,\bSigma_{11})$ depend on $(\alpha,\beta,\gamma)$ per \eqref{eq:kernel}.

\begin{remark}\label{re:whyGP}
The mapping $y(\btheta)$ is deterministic: Given $\btheta$, the setpoint $y(\btheta)$ can be found by solving \eqref{eq:OPF2} for the requested $\btheta$. Of course, if $\btheta$ is random, then $y(\btheta)$ becomes random. Many problems in grid operations rely on approximating the PDF or estimating statistics (mean or covariance) of $y(\btheta)$ given the PDF or samples of $\btheta$; see \cite{TJKT20} and references therein. This work does not consider the aforesaid problem. Here $\btheta$ and $y(\btheta)$ are both deterministic. What is modeled as a GP is our estimate $\hat{y}(\btheta)$ of $y(\btheta)$ given dataset $\mcT$. It is our model estimates for given and future data that are modeled as jointly Gaussian in \eqref{eq:GP}. This agrees with the principle of Bayesian inference wherein an unknown quantity $y$ is assigned a prior PDF even if $y$ is deterministic. Upon collecting data related to $y$, one computes the posterior PDF of $y$. The final estimate $\hat{y}$ for $y$ is exactly the mean of the posterior distribution. For example, GP modeling has been widely used in geostatistics where scientists would like to build a topographical map of an area using elevation readings collected at a finite number of locations. To be able to interpolate from the collected elevation samples to unobserved points, a GP model is postulated even though elevation is a deterministic process, not a random one. If $y(\btheta)$ is relatively smooth, a kernel function such as the Gaussian one in \eqref{eq:kernel} can capture the variation of $y$ over $\btheta$'s. The analogy carries over to the OPF problem at hand. 
\end{remark}

\begin{remark}\label{re:joint}
We decided to build a separate GP model for each entry of the OPF minimizer $\bx_{\btheta}$, or at least for the entries of $\bx_{\btheta}$ we are interested in. Alternatively, we could pursue training GP models jointly over all inverters. In this case, one should come up with meaningful kernels over $\btheta$'s and buses alike, i.e., $k\left((\btheta_i,n),(\btheta_j,m)\right)$, to jointly train GP models for $x_n(\btheta)$ and $x_m(\btheta)$. To simplify the joint kernel design task, a product structure is oftentimes imposed according to which $k\left((\btheta_i,n),(\btheta_j,m)\right)=k_1\left(\btheta_i,\btheta_j\right)\cdot k_2(n,m)$; see e.g.,~\cite{GP4dynamicsPRWRS21}, \cite{KZG14}. Whether such structure makes sense for GP-OPF and how to parameterize $k_2(n,m)$ in a physics-informed manner are non-trivial but relevant questions, which go beyond the scope of this work.
\end{remark}

\section{Sensitivity-Informed GP-OPF (SI-GP-OPF)}\label{sec:SI-GP-OPF}
So far, the training dataset $\mcT$ consists of pairs of OPF parameters and solutions. This complies with the standard supervised learning setup. Nonetheless, when one aims at predicting solutions to a parametric optimization, there is more information to be exploited. Incorporating such rich information of OPF solutions can improve data efficiency in the sense that: \emph{i)} A learner could infer $y(\btheta)$ with the same estimation accuracy using a smaller training dataset $\mcT$. This is computationally advantageous as fewer instances of \eqref{eq:OPF} have to be solved; or \emph{ii)} A learner could achieve higher estimation accuracy for the same $\mcT$. We next elaborate on one of the additional information on $y(\btheta)$ the learner can exploit.

When it comes to a parametric problem [cf.~\eqref{eq:OPF2}], one can compute the partial derivatives of the minimizer $\bx_{\btheta}$ with respect to parameters $\btheta$ using sensitivity analysis; see~\cite{fiacco1976sensitivity}, \cite{Conejo06}. Given the primal/dual solution of an optimization, computing the Jacobian matrix $\nabla_{\btheta}\bx_{\btheta}$ carrying the sensitivities of $\bx_{\btheta}$ with respect to $\btheta$ is as easy as solving a system of linear equations as long as the problem involves continuously differentiable objective and constraint functions. We defer the sensitivity analysis of \eqref{eq:OPF2} to Sec.~\ref{sec:sa}. For now, let us suppose that along with the solution $y(\btheta_t)$ for each $t$, the learner has also computed the $M$-length gradient $\nabla_{\btheta} y(\btheta_t)$, denoted by $\dot{\by}(\btheta_t)$ for short. We thus augment the training dataset as
\begin{equation}\label{eq:augment}
\mcT=\{\left(\btheta_t,y(\btheta_t)\right)\}_{t=1}^T \rightarrow \bar{\mcT}=\{\left(\btheta_t,y(\btheta_t),\dot{\by}(\btheta_t)\right)\}_{t=1}^T.
\end{equation}
The new dataset $\bar{\mcT}$ carries $(M+1)$ pieces of information per OPF example rather than just one as in the original $\mcT$. In other words, for each sampled $\btheta_t$, we now know not only the value of the OPF mapping $y(\btheta)$, but also its gradient $\dot{y}(\btheta)$. The gradient information $\dot{y}(\btheta)$ is important as along with $y(\btheta)$, it approximates the mapping $y(\btheta)$ in a neighborhood around $\btheta_t$ through a first-order Taylor's series expansion. Moreover, knowing the gradients, we can train a machine learning that given a new $\btheta$, predicts both $y(\btheta)$ and its gradient. 

The pertinent question now is whether the extra information of gradients can be incorporated into GP-OPF. Can the additional data $\{\dot{\by}(\btheta_t)\}_{t=1}^T$ be included in the GP model as part of vector $\by_1$ in \eqref{eq:GP}? The requirement for achieving this is that gradient data can also be modeled as GPs, and that their covariances (appearing as blocks of $\bSigma_{11}$ and $\bSigma_{12}$ in \eqref{eq:GP}) can be described by a parametric model. Fortunately, an appealing property of GPs is that the derivative of a GP with respect to its independent variable ($\btheta$ in our case) is a GP itself.~
In particular, if $y(\btheta)$ is a zero-mean GP, the gradient $\dot{\by}(\btheta)$ is a zero-mean GP as well. Additionally, the covariance between $y(\btheta)$ and $\dot{\by}(\btheta)$ can be derived as
\begin{align}\label{eq:Kgrad}
\mathbb{E}[y(\btheta_i)\dot{\by}(\btheta_j)]
&=\mathbb{E}[y(\btheta_i)\nabla_{\btheta_j}y(\btheta_j)]\nonumber\\
&=\nabla_{\btheta_j}\mathbb{E}[y(\btheta_i)y(\btheta_j)]\nonumber\\
&=\nabla_{\btheta_j}k(\btheta_i,\btheta_j)
\end{align}
where the second equality follows by exchanging the order of expectation and differentiation, and the third equality is by definition of the kernel function. The covariance between gradient vectors can be derived similarly as
\begin{equation*}
\mathbb{E}[\dot{\by}(\btheta_i)\dot{\by}^\top(\btheta_j)]=\nabla_{\btheta_i\btheta_j}^2\mathbb{E}[y(\btheta_i)y(\btheta_j)]=\nabla_{\btheta_i\btheta_j}^2k(\btheta_i,\btheta_j).
\end{equation*}

For the Gaussian kernel in \eqref{eq:kernel}, the aforesaid quantities can be readily computed as
\begin{align*}
\nabla_{\btheta_j}k(\btheta_i,\btheta_j)&=\beta k(\btheta_i,\btheta_j)\left(\btheta_i-\btheta_j\right)\\
\nabla_{\btheta_i\btheta_j}^2k(\btheta_i,\btheta_j)&=
\beta k(\btheta_i,\btheta_j) [ \bI_M-\beta\left(\btheta_i-\btheta_j\right)\left(\btheta_i-\btheta_j\right)^\top ]
\end{align*}
where $\bI_M$ is the identity matrix of size $M$. As in \eqref{eq:kernel+noise}, gradient labels are observed under modeling noise so for some $\epsilon>0$: \[\mathbb{E}[\dot{\by}(\btheta_i)\dot{\by}(\btheta_j)^\top]=\nabla_{\btheta_i\btheta_j}^2k(\btheta_i,\btheta_j)+\epsilon\delta_{ij}\bI_M.\]

Since gradients comply with the GP model, they can be included in the GP framework presented earlier. Stacking this extra information modifies $\by_1$ of \eqref{eq:GP} from a $T$-length vector carrying only $y(\btheta_t)$'s to a vector of dimension $\bar{T}:=T(M+1)$:
\begin{equation*}
 \bby_1^\top := \left[\by_1^\top~~\dot{\by}^\top(\btheta_1)~~\cdots~~\dot{\by}^\top(\btheta_T)\right].
\end{equation*}
Vector $\bby_1$ replaces $\by_1$ in \eqref{eq:GP} and \eqref{eq:mmse}. Mean vectors remain zero and the covariances can be computed as explained earlier. It is worth noting that incorporating sensitivities into GP models is quite standard~\cite[Sec.~9.4]{GP}. The novelty here is that the sensitivities are indeed available for optimization data and can be obtained with minimal computational overhead. Deferring the computation of $\dot{\by}(\btheta)$ to Section~\ref{sec:sa}, we next address the computational issues arising when expanding data by $(M+1)$ times while migrating from $\by_1$ to $\bby_1$.

\section{Random Feature-Based SI-GP-OPF}\label{sec:rf}
GP-OPF suffers from the curse of dimensionality~\cite{GP}. If the size of the training dataset is $T$, inverting matrix $\bSigma_{11}$ in \eqref{eq:mmse} takes $\mcO(T^3)$ operations during training. With $\bSigma_{11}^{-1}$ and $\bSigma_{11}^{-1}\by_1$ computed, prediction takes $\mcO(T)$ for the mean in \eqref{eq:mmse:m}, and $\mcO(T^2)$ for the variance in \eqref{eq:mmse:c} per new test case. To render GP learning scalable, existing solutions include low-rank~\cite{SparseGP}, structured approximants of $\bSigma_{11}$~\cite{Kernelsampling}, and random features~\cite{MKL_RF}. The scaling issue of GPs is exacerbated with SI-GPs. Albeit gradient labels are introduced to reduce the number of OPF instances $T$ that need to be solved, they increase the size of the augmented dataset as $\bar{T}=(M+1)T$. This section extends the concept of \emph{random features (RFs)} to gradient data to enable scalable learning of the OPF mapping. 

We first present the plain RF-based GP-OPF (RF-GP-OPF). The crux in GP-OPF is inverting $\bSigma_{11}$. The idea of random features is to approximate the kernel function of \eqref{eq:kernel} as the inner product between two $D$-length vectors~\cite{Recht_RF}
\begin{equation}\label{eq:kernel_approx}
    k(\btheta_i,\btheta_j) \simeq \alpha \bz^\top(\btheta_i)\bz(\btheta_j)
\end{equation}
where vector $\bz(\btheta)$ is a randomized nonlinear (sinusoidal) transformation of $\btheta$. Its $d$-th entry is defined as
\begin{equation}\label{eq:zd}
    z_d(\btheta):=\sqrt{\tfrac{2}{D}}\cos\left(\bv_d^\top\btheta + \phi_d\right),\quad d=1,\ldots,D.
\end{equation}
Here $\{\bv_d\}_{d=1}^D$ are random vectors drawn independently from $\mcN(\bzero,\beta\bI_M)$, and $\{\phi_d\}_{d=1}^D$ are random scalars drawn uniformly from $[0,2\pi]$. Recall $\beta$ is one of the parameters of the Gaussian kernel in \eqref{eq:kernel} and $M$ is the length of $\btheta$. Vector $\bz(\btheta_i)$ constitutes the vector of \emph{random features} that transforms datum $\btheta_i\in\mathbb{R}^M$ to $\bz(\btheta_i)\in\mathbb{R}^D$. For an explanation of why \eqref{eq:kernel_approx} is a valid approximation and how $D$ affects accuracy, the interested reader is referred to Appendix~\ref{sec:AppA}. Reference~\cite{Recht_RF} shows that \eqref{eq:kernel_approx} approximates $k(\btheta_i,\btheta_j)$ within $\epsilon$ uniformly over $(\btheta_i,\btheta_j)$ if $D$ is selected as $\mcO(M\epsilon^{-2}\log\epsilon^{-2})$, though excellent regression results are empirically observed with smaller $D$.

Let us collect the RF vectors for all training data $\{\bz(\btheta_t)\}_{t=1}^T$ as rows of a $T\times D$ matrix $\bZ_1$. Due to \eqref{eq:kernel_approx}, we can approximate the covariance $\bSigma_{11}$ from \eqref{eq:kernel}--\eqref{eq:kernel+noise} as
\begin{equation}\label{eq:hSigma11}
\hbSigma_{11}= \alpha \bZ_1\bZ_1^\top + \gamma \bI_T.
\end{equation}
Similarly to $\bZ_1$, let the $S\times D$ matrix $\bZ_2$ collect the RF vectors for all data of the validation dataset $\mcS$. Then, the cross-covariance $\bSigma_{21}$ can be approximated as
\begin{equation}\label{eq:hSigma21}
\hbSigma_{21}= \alpha \bZ_2\bZ_1^\top.
\end{equation}
From \eqref{eq:hSigma11}--\eqref{eq:hSigma21}, we can approximate $\bSigma_{21}\bSigma_{11}^{-1}$ as 
\begin{subequations}\label{eq:RFadvantage}
\begin{align}
\hbSigma_{21}\hbSigma_{11}^{-1}&=\alpha \bZ_2\bZ_1^\top\left(\alpha \bZ_1\bZ_1^\top + \gamma \bI_T\right)^{-1}\label{eq:RFadvantage:a}\\
&=\alpha \bZ_2\left(\alpha \bZ_1^\top\bZ_1 + \gamma \bI_D\right)^{-1}\bZ_1^\top\label{eq:RFadvantage:b}
\end{align}   
\end{subequations}
where the second equality follows from the matrix inversion lemma. The key point is that \eqref{eq:RFadvantage:b} involves inverting a $D\times D$ rather than a $T\times T$ matrix as in \eqref{eq:RFadvantage:a}. Leveraging~\eqref{eq:RFadvantage}, Table~\ref{tbl:steps} reports the steps of RF-GP-OPF and their complexity. If $T>D>M$, the training phase takes $\mcO(D^2T)$, while the prediction phase costs $\mcO(DM)$ for the mean and $\mcO(D^2)$ for the variance per new datum. If $T<M$, RF-GP-OPF has no advantage over the plain GP-OPF.

\begin{table}[t]
\renewcommand{\arraystretch}{1.1}
    \caption{Training/Prediction with RF-GP-OPF}
  \label{tbl:steps}
  \centering\normalsize
\begin{tabular}{@{}p{6.7cm}r} 
  \hline
    \emph{Training Phase} & $\mcO(D^2T)$\\ 
    \hline
    \emph{T1)} Draw $\{\bv_d,\phi_d\}_{d=1}^D$ and find $\bZ_1$ from \eqref{eq:zd}            &$\mcO(DTM)$  \\
    \emph{T2)} Compute $\bZ_1^\top\bZ_1$ & $\mcO(D^2T)$\\
    \emph{T3)} Compute $\bZ_1^\top\by_1$ & $\mcO(DT)$\\
    \emph{T4)} Invert $\alpha \bZ_1^\top\bZ_1 + \gamma \bI_D$ & $\mcO(D^3)$\\
    \emph{T5)} Compute $(\alpha \bZ_1^\top\bZ_1 + \gamma \bI_D)^{-1}(\bZ_1^\top\by_1)$          & $\mcO(D^2)$\\
    \emph{T6)} Compute $(\alpha \bZ_1^\top\bZ_1 + \gamma \bI_D)^{-1}(\bZ_1^\top\bZ_1)$ & $\mcO(D^3)$\\
    \hline    
    \emph{Prediction Phase (per new OPF instance)}\\
    \hline
    \emph{P1)} Compute $\bZ_2\in\mathbb{R}^{1\times D}$ from \eqref{eq:zd} & $\mcO(DM)$\\
    \emph{P2)} Premultiply the result of \emph{T5)} by $\alpha\bZ_2$ to find predictive mean \eqref{eq:mmse:m} & $\mcO(D)$\\
    \emph{P3)} Pre/post-multiply the result of \emph{T6)} by $\alpha\bZ_2$ ($\alpha\bZ_2^\top$) to find predictive variance \eqref{eq:mmse:c}  & $\mcO(D^2)$\\
    \hline    
  \end{tabular}
\end{table}



An RF-based implementation is really helpful when migrating from GP-OPF to SI-GP-OPF as now the size of the training dataset explodes from $T$ to $\bar{T}=T(M+1)$. If one implements SI-GP-OPF using RFs (hereafter termed RF-SI-GP-OPF) shortsightedly as per Table~\ref{tbl:steps}, the complexity over training would be $\mcO(D^2TM)$. A smarter implementation exploiting the problem structure can reduce the complexity further to $\mcO(DTM+D^3+D^2(T+M))$ as explained next.

We are interested in finding unbiased estimates of $\mathbb{E}\left[ y(\btheta_i) \dot{\by}(\btheta_j)\right]$ and $\mathbb{E}[\dot{\by}(\btheta_i)\dot{\by}^\top(\btheta_j)]$. We rely on \eqref{eq:Kgrad}. If \eqref{eq:kernel_approx} is an estimate of $k(\btheta_i,\btheta_j)$, then $\nabla_{\btheta_j} k(\btheta_i,\btheta_j)$ and $\nabla_{\btheta_i\btheta_j}^2k(\btheta_i,\btheta_j)$ can be approximated by inner products as
\begin{subequations}\label{eq:si-rf}
    \begin{align}\label{eq:zdot}
         \nabla_{\btheta_j} k(\btheta_i,\btheta_j) &\simeq\alpha \left(\nabla_{\btheta_j} \bz(\btheta_j)\right)^\top\bz(\btheta_i) \\\label{eq:zddot}
         \nabla_{\btheta_i\btheta_j}^2k(\btheta_i,\btheta_j) &\simeq\alpha
         \left(\nabla_{\btheta_i} \bz(\btheta_i)\right)^\top \nabla_{\btheta_j} \bz(\btheta_j)
    \end{align}
\end{subequations}
where $\nabla_{\btheta_j} \bz(\btheta_j)$ is a $D\times M$ Jacobian matrix. The $d$-th row of this Jacobian can be computed by differentiating \eqref{eq:zd} to get
\begin{equation*}
    \nabla_{\btheta_j}z_d(\btheta_j) = -\sqrt{\tfrac{2}{D}} \sin(\bv_d^\top \btheta_j+\phi_d) \bv_d.
\end{equation*}

\begin{table}[t]
\renewcommand{\arraystretch}{1.1}
    \caption{Training/Prediction with RF-SI-GP-OPF}
  \label{tbl:steps2}
  \centering\normalsize
\begin{tabular}{@{}p{6.1cm}r}   
  \hline
    \emph{Training} $\mcO(DTM+D^2(D+T+M))$\\ 
    \hline
    \emph{T1)} Draw $\{\bv_d,\phi_d\}_{d=1}^D$ and find $\bbZ_1$       &$\mcO(DTM)$\\
    \emph{T2)} Find $\bbZ_1^\top\bD^{-1}\bbZ_1$ & $\mcO(D^2(T{+}M))$\\
    \emph{T3)} Find $\bbZ_1^\top\bD^{-1}\bby_1$ & $\mcO(DTM)$\\
    \emph{T4)} Invert $\bbZ_1^\top \bD^{-1}  \bbZ_1 +\bI_D$ & $\mcO(D^3)$\\
    \emph{T5)} Find $(\bbZ_1^\top \bD^{-1}  \bbZ_1 +\bI_D)^{-1}(\bbZ_1^\top\bD^{-1}\bby_1)$          & $\mcO(D^2)$\\
    \emph{T6)} Find $(\bbZ_1^\top \bD^{-1}  \bbZ_1 +\bI_D)^{-1}(\bbZ_1^\top\bD^{-1}\bbZ_1)$ & $\mcO(D^3)$\\
    \hline    
    \emph{Prediction (as in Table~\ref{tbl:steps})}\\
    \hline    
  \end{tabular}
\end{table}

For the computational advantage shown in~\eqref{eq:RFadvantage} to carry over to SI-GPs, we should be able to express the covariance $\mathbb{E}\left[\bar{\by}_1 \bar{\by}_1^\top \right]$ as a rank--$D$ plus a scaled identity matrix. To this end, define the quantities
\begin{equation*}
s_d(\btheta):=-\sqrt{\tfrac{2}{D}} \sin(\bv_d^\top \btheta+\phi_d), \quad d=1,\dots,D
\end{equation*}
and stack them in vector $\bs(\btheta_j):=\left[s_1(\btheta_j)~\cdots~s_D(\btheta_j)\right]^\top$. It is not hard to verify that the Jacobian matrix $\nabla_{\btheta_j} \bz(\btheta_j)$ can be expressed as the Khatri-Rao product 
\begin{equation*}
     \nabla_{\btheta_j}\bz(\btheta_j)=\left(\bs^\top(\btheta_j) \ast \bV \right)^\top
\end{equation*}
where $\bV:=[\bv_1~\cdots~\bv_D]$ is a $M\times D$ matrix. If we place vectors $\{\bs(\btheta_t)\}_{t=1}^T$ as rows of matrix $\bS_1$, we can approximate
\begin{align}\label{eq:cov_ybar}
\mathbb{E}[\bar{\by}_1 \bar{\by}_1^\top ]&\simeq \alpha  \bbZ_1 \bbZ_1^\top + \bD\\
\text{where}~~\bbZ_1&:=
    \begin{bmatrix}
    \bZ_1\\
    \bS_1 \ast \bV
    \end{bmatrix}
    ~~\text{and}~~
    \bD:=\begin{bmatrix}
    \gamma \bI_T & \bzero\\
    \bzero & \epsilon \bI_{MT}
    \end{bmatrix}.\nonumber
\end{align}
Since $\bZ_1$ and $\bS_1$ are $T\times D$ and $\bV$ is $M\times D$, matrix $\bbZ_1$ is $T(M+1)\times D$. Thanks to~\eqref{eq:cov_ybar}, the computational advantage of~\eqref{eq:RFadvantage} carries over to SI-GPs as now $\bSigma_{11}=\mathbb{E}[\bar{\by}_1 \bar{\by}_1^\top]$ and
\begin{subequations}
    \begin{align*}
    \hat{\bSigma}_{21} \hat{\bSigma}_{11}^{-1} &= \alpha \bZ_2 \bbZ_1^\top \left ( \alpha \bbZ_1\bbZ_1^\top + \bD \right)^{-1}\\
    &=\alpha \bZ_2 \left(\alpha \bbZ_1^\top \bD^{-1}  \bbZ_1 +\bI_D \right)^{-1}  \bbZ_1^\top \bD^{-1}.
    \end{align*}
\end{subequations}
Again, we need to invert a $D\times D$ instead of a $\bar{T} \times \bar{T}$ matrix. Table~\ref{tbl:steps2} details the computational complexity per step of RF-SI-GP-OPF. Interestingly, steps \emph{T1)}--\emph{T2)} of Table~\ref{tbl:steps2} maintain the complexity of steps \emph{T1)}--\emph{T2)} of Table~\ref{tbl:steps} despite $T$ has been replaced by $\bar{T}$: Regarding \emph{T1)}, a careful yet mundane analysis on \eqref{eq:cov_ybar} shows that $\bbZ_1$ can indeed be computed in $\mcO(DTM)$, and not $\mcO(D\bar{T}M)$. Step \emph{T2)} takes $\mcO(D^2T)$ as the properties of the Khatri-Rao product yield
\begin{align*}
\bbZ_1^\top \bD^{-1}  \bbZ_1 &=\gamma^{-1}\bZ_1^\top\bZ_1 +\epsilon^{-1}(\bS_1\ast\bV)^\top (\bS_1\ast\bV)\\
&=\gamma^{-1}\bZ_1^\top\bZ_1 +\epsilon^{-1}(\bS_1^\top\bS_1)\circ(\bV^\top\bV)
\end{align*}
where $\circ$ denotes the Hadamard (entry-wise) matrix multiplication. Evidently, the complexity of RF-SI-GP-OPF is of the same order as that of RF-GP-OPF. 

{In summary, the suggested RF-SI-GP-OPF is implemented as described next. Data generation step \emph{s1)} is shared across all inverters. This involves solving $T$ OPF instances and computing sensitivities. Learning step \emph{s2)} is implemented separately per inverter. It includes finding kernel hyperparameters via MLE and following the steps in Table~\ref{tbl:steps2}. Both \emph{s1)} and \emph{s2)} occur offline. In real time, setpoints are computed per inverter according to the prediction phase of Table~\ref{tbl:pred_times}. This phase entails only $\mcO(DM)$ or $\mcO(D^2)$ calculations depending on whether predictive variance is needed or not. Because $D>T>M$ in our tests, the computational time during real-time operation is independent of $M$ and scales linearly with the number of inverters. If we assume that $M$ and $D$ scale linearly with $N$, and that $N_g=N$, the prediction phase has a worst-case complexity of $\mcO(N^3)$. This still outperforms interior point-based OPF solvers (LP, QP, SOCP) whose complexities are $\mcO(N^\kappa)$ with $\kappa > 3.5$.}

\section{Sensitivity Analysis of SOCP-based OPF}\label{sec:sa}
Section~\ref{sec:GP-OPF} expanded dataset $\mcT$ to $\bar{\mcT}$ by including the gradient vector $\nabla_{\btheta} y(\btheta)$ and recall $y(\btheta)$ denotes one of the entries of the minimizer $\bx(\btheta)$ of \eqref{eq:OPF2}. This section performs sensitivity analysis of \eqref{eq:OPF2} to compute the Jacobian matrix of the entire vector $\bx(\btheta)$ with respect to $\btheta$. Once presented with a $\btheta$, an off-the-shelf SOCP solver can be used to obtain the optimal primal/dual variables of~\eqref{eq:OPF2}. Dropping the subscript $\btheta$, let us denote the minimizer as $\bx$, and the optimal dual variables for \eqref{eq:OPF2:eq}--\eqref{eq:OPF2:soc} as $\blambda$, $\bmu$, and $\{\bar{\bnu}_m\}_{m=1}^M$, respectively. The lengths of $\blambda$ and $\bmu$ coincide with the number of equality and inequality constraints, while $\bar{\bnu}_m$'s are the dual conic variables with lengths one more than the number of rows in $\bA_m$. The dual variables obtained from most SOCP solvers correspond to the conic approach-based Lagrangian function
\begin{align*}
\bc^\top\bx&+\blambda^\top(\bA_e \bx- \bB_e \btheta - \bef_e)+\bmu^\top(\bA_i \bx - \bB_i \btheta - \bef_i)\\
&+\sum_{m=1}^{2N} \bar{\bnu}_m^\top
\left[\begin{array}{c}
-(\bb_m^\top\bx+f_m)\\
\bA_m\bx
\end{array}\right].
\end{align*}

Deviating from the conic-approach, for sensitivity computation we invoke the direct approach-based Lagrangian where the SOCs in~\eqref{eq:OPF2:soc} are treated as scalar constraints
\begin{align*}
\mcL(\bx,\blambda,\bmu, \bnu;\btheta)&=\bc^\top\bx+\blambda^\top(\bA_e \bx- \bB_e \btheta - \bef_e)\\
&~~+\bmu^\top(\bA_i \bx - \bB_i \btheta - \bef_i)\\
&~~+\sum_{m=1}^{2N} \nu_m\left(\|\bA_m \bx\| - \bb_m^\top \bx  - f_m\right).
\end{align*}
It can be shown that the optimal dual variables obtained from the conic approach satisfy the Karush–Kuhn–Tucker (KKT) conditions for the direct approach with $\nu_m$ being equal to the first entry of $\bbnu_m$~\cite{SOCP_Lobo}. This allows us to obtain the optimal dual variables from a conic solver and proceed with sensitivity analysis using the KKT conditions for the direct approach. 

Sensitivity analysis aims at finding infinitesimal changes $(\d\bx,\d\blambda,\d\bmu,\d\bnu)$, so that the perturbed point $(\bx+\d\bx, \blambda+\d\blambda,\bmu+\d\bmu, \bnu+\d\bnu)$ satisfies the first-order KKT conditions when the problem parameters change from $\btheta$ to $\btheta+\d\btheta$~\cite{fiacco1976sensitivity}. To this end, we first review optimality conditions and then apply implicit differentiation to compute the sought sensitivities. Starting with Lagrangian optimality condition $\nabla_{\bx}\mcL=\bzero$:
\begin{equation}\label{eq:kktLagrange}
  \bc+\bA_e^\top\blambda+\bA_i^\top\bmu+\sum_{m=1}^{2N}\nu_m\left(\frac{\bA_m^\top\bA_m\bx}{\|\bA_m \bx\|}-\bb_m\right)=\bzero
\end{equation}
The first-order optimality conditions further include primal feasibility~\eqref{eq:OPF2:eq}--\eqref{eq:OPF2:soc}; dual feasibility $\bmu\geq \bzero$ and $\bnu\geq\bzero$; and complementary slackness
\begin{subequations}\label{eq:kktcs}
    \begin{align}
    \diag(\bmu)\left(\bA_i \bx - \bB_i \btheta - \bef_i\right)&=\bzero\label{seq:kktcs:a}\\
    \nu_m\left(\|\bA_m \bx\| - \bb_m^\top \bx  - f_m\right)&=0,\quad \forall m.\label{seq:kktcs:b}
    \end{align}
\end{subequations}
We next compute the total differentials over $(\d\bx,\d\blambda,\d\bmu,\d\bnu,\d\btheta)$ for the first-order optimality conditions involving equalities, namely conditions~\eqref{eq:OPF2:eq}, \eqref{eq:kktLagrange}, and \eqref{eq:kktcs}:
\begin{subequations}\label{eq:td}
    \begin{align}
    &\bA_e\d\bx-\bB_e\d\btheta=\bzero\\
    &\bA_e^\top\d\blambda+\bA_i^\top\d\bmu+\sum_{m=1}^{2N}\left(\frac{\bA_m^\top\bA_m\bx}{\|\bA_m \bx\|}-\bb_m\right)\d\nu_m\notag\\
    &+\sum_{m=1}^{2N}\nu_m\left(\frac{\bA_m^\top\bA_m}{\|\bA_m \bx\|}-\frac{\bA_m^\top\bA_m\bx\bx^\top\bA_m^\top\bA_m}{\|\bA_m \bx\|^3}\right)\d\bx=\bzero\\
    &\diag(\bmu)\left(\bA_i \d\bx - \bB_i \d\btheta\right)+\diag\left(\bA_i \bx - \bB_i \btheta - \bef_i\right)\d\bmu=\bzero\label{seq:td:c}\\
    &\nu_m\left(\frac{\bx^\top\bA_m^\top\bA_m}{\|\bA_m \bx\|}-\bb_m^\top\right)\d\bx\notag\\ 
    &\quad\quad\quad\quad+\left(\|\bA_m \bx\| - \bb_m^\top \bx  - f_m\right)\d\nu_m=0,~\forall m.
    \end{align}
\end{subequations}

Given an optimal primal/dual solution $(\bx,\blambda,\bmu,\bnu)$ of~\eqref{eq:OPF2}, a perturbed point $(\bx+\d\bx, \blambda+\d\blambda,\bmu+\d\bmu, \bnu+\d\bnu)$ satisfies the KKT equality conditions \eqref{eq:OPF2:eq}, \eqref{eq:kktLagrange}, and \eqref{eq:kktcs} for $\btheta+\d\btheta$, if the conditions in \eqref{eq:td} are satisfied. Interestingly, it can be shown that under \emph{strict complementary slackness}, satisfying~\eqref{eq:td} ensures that the perturbed point satisfies the inequality conditions~\eqref{eq:OPF2:ineq}--\eqref{eq:OPF2:soc}, and the dual feasibility conditions as well~\cite{L2O2021}. In other words, although the perturbed point was constructed by taking into account only the optimality conditions expressed as equalities, it also satisfies the optimality conditions expressed as inequalities. Hence, the perturbed point constitutes an optimal primal/dual solution of~\eqref{eq:OPF2} for parameter $\btheta+\d\btheta$.\footnote{Strict complementary slackness means the optimal dual variables corresponding to binding inequality constraints are strictly positive. While this condition generally holds numerically, in the advent of a degenerate scenario violating complementary slackness, such samples can be removed from the training dataset or included without their sensitivities.} Defining $\bdelta:=[\bx^\top~\blambda^\top~\bmu^\top~\bnu^\top]^\top$, the system of equations in~\eqref{eq:td} can be compactly represented as
\begin{equation}\label{eq:SU}
    \bS\d\bdelta=\bU\d\btheta
\end{equation}
where $(\bS,\bU)$ follow from~\eqref{eq:td} and depend on $(\bx,\blambda,\bmu,\bnu)$ as detailed in Appendix~\ref{sec:AppB}. 

If $\bS$ is invertible, the desired sensitivities $\nabla_{\btheta}\bx$ can be extracted as the top rows of $\bS^{-1}\bU$ corresponding to $\bx$. However, OPF instances leading to singular $\bS$ have been reported for transmission and distribution systems~\cite{SGKCB2020}, \cite{Dorfler18LICQ}. Investigating further into such scenarios, reference~\cite{L2O2021} argues that the invertibility of $\bS$ is not necessarily needed to compute $\nabla_{\btheta}\bx$ using~\eqref{eq:SU}. Rather, given a vector $\d\btheta$, if~\eqref{eq:SU} has a unique solution for $\d\bx$, the sensitivities $\nabla_{\btheta}\bx$ do exist. Uniqueness in $\d\bx$ is guaranteed if for any vector in $\nullspace(\bS)$, its top entries corresponding to $\bx$ are zero. Therefore, if the basis of $\nullspace(\bS)$ exhibits the desired sparsity, then $\nabla_{\btheta}\bx$ exists. If it exists, matrix $\nabla_{\btheta}\bx$ can be extracted as the top rows of $\bS^\dagger\bU$ corresponding to $\bx$, where $\bS^\dagger$ is the pseudoinverse of $\bS$. Ultimately, existence and computation of $\nabla_{\btheta}\bx$ are ensured by two technical assumptions: strict complementary slackness and a second-order optimality condition; see~\cite{L2O2021} for details. In this work, the steps followed to augment the training set $\mcT$ with sensitivities $\nabla_{\btheta}\bx$ are: \emph{i)} Use the optimal solution $(\bx,\blambda,\bmu,\bnu)$ obtained from an SOC solver to construct the system of linear equations in~\eqref{eq:SU}; \emph{ii)} Numerically verify the existence of $\nabla_{\btheta}\bx$ by checking the sparsity pattern for the basis of $\nullspace(\bS)$; and \emph{iii)} Extract $\nabla_{\btheta}\bx$ as the appropriate rows of $\bS^\dagger\bU$.

\begin{remark}\label{re:diff}
Our analysis presumed the cost and constraint functions of \eqref{eq:OPF2} are differentiable with respect to $\bx$ and $\btheta$. This is true with the exception of the SOCs in \eqref{eq:OPF2:soc}. The constraint function $\|\bA_m\bx\|-\bb_m^\top\bx -f_m$ becomes non-differentiable if $\bA_m\bx=\bzero$ at optimality. For the SOCs corresponding to the convex relaxation in \eqref{eq:OPF:soc}, this problematic scenario cannot occur because if \eqref{eq:OPF2} is feasible and the relaxation is exact, then $\|\bA_m\bx\|=v_{\pi_n}+\ell_n\geq \underline{v}_{\pi_n}>0$. 

The apparent power constraints in~\eqref{eq:OPF:smax} call for a more careful treatment. Suppose the $m$-th apparent power constraint corresponds to inverter $n$. When posed as the SOC $\|\bA_m\bx\|\leq f_m$ with $f_m=\bar{s}_n^g>0$, the constraint function becomes non-differentiable if $\|\bA_m\bx\|=0$, or equivalently, $p_n^g=q_n^g=0$ at optimality. Complementary slackness yields also $\nu_m=0$. To deal with any non-differentiable constraint, we will perform sensitivity analysis assuming all troublesome constraints have been replaced by their original convex quadratic form in \eqref{eq:OPF:smax}. That form is differentiable and thus amenable to sensitivity analysis. Let $\nu_m'$ be the optimal dual for the differentiable form of the constraint. The primal/dual solutions of the two OPF formulations coincide and $\nu_m'=\nu_m=0$. Standard results from sensitivity analysis show that if a constraint is non-binding under $\btheta$, it remains non-binding under any $\btheta+\d \btheta$ and so $\nu_m'=\d\nu_m'=0$; see~\cite{Conejo06}. Because of this, the troublesome constraint can be ignored when forming \eqref{eq:SU}. While the aforementioned discussion serves well for completeness, our numerical tests did not encounter instances of $\|\bA_m\bx\|=0$.
\end{remark}

\section{Numerical Tests}\label{sec:tests}
GP-OPF was tested on the IEEE 13- and 123-bus benchmarks converted to single-phase~\cite{jalaliSVM}. Real-world minute-based active load and solar generation data from 446 households were obtained from the Smart* Project collected on July 1 of 2011 and 2015, respectively~\cite{smart_load}. Load demand per bus was simulated by adding up 20 randomly sampled household demands, and scaling this aggregate demand so its maximum matches the benchmark value for that bus. Sequences of solar generation were obtained upon aggregation and scaling similarly. Solar sequences were scaled on a per feeder and bus basis to simulate different solar penetration levels as described later. Lacking reactive power data, we simulated lagging power factors uniformly and independently drawn from $[0.9,1.0]$ across buses, which were kept fixed across time. To allow for reactive power injection at maximum solar irradiance, we oversized inverters by $10\%$. All tests were ran on a 2.9~GHz AMD 7-core processor laptop computer with 16~GB RAM. 

The labeled dataset was generated by solving the OPF in~\eqref{eq:OPF} using the YALMIP toolbox and the SOCP solver SDPT3. We used the GPML toolbox for estimating $(\alpha, \beta, \gamma)$. Having estimated these parameters, we subsequently estimated $\epsilon$ using MATLAB's \texttt{fitrgp} function. This toolbox accepts custom covariance functions, which is needed for introducing the covariance of sensitivities and finding $\epsilon$. Covariance matrices were formed using the learned hyper-parameters per~\eqref{eq:Kgrad}, and Tables~\ref{tbl:steps} and~\ref{tbl:steps2}. The prediction accuracy of a GP-OPF estimate $\hat{x}_n$ of $x_n$ was measured using the relative percent error (RPE) which is useful when $x_n$ can take zero values $\textrm{RPE}_n:= |\hat{x}_n-x_n|/\left(\frac{1}{N}\sum_{m=1}^Nx_m\right)$.


\begin{figure}[t]
\centering
\includegraphics[scale=0.26]{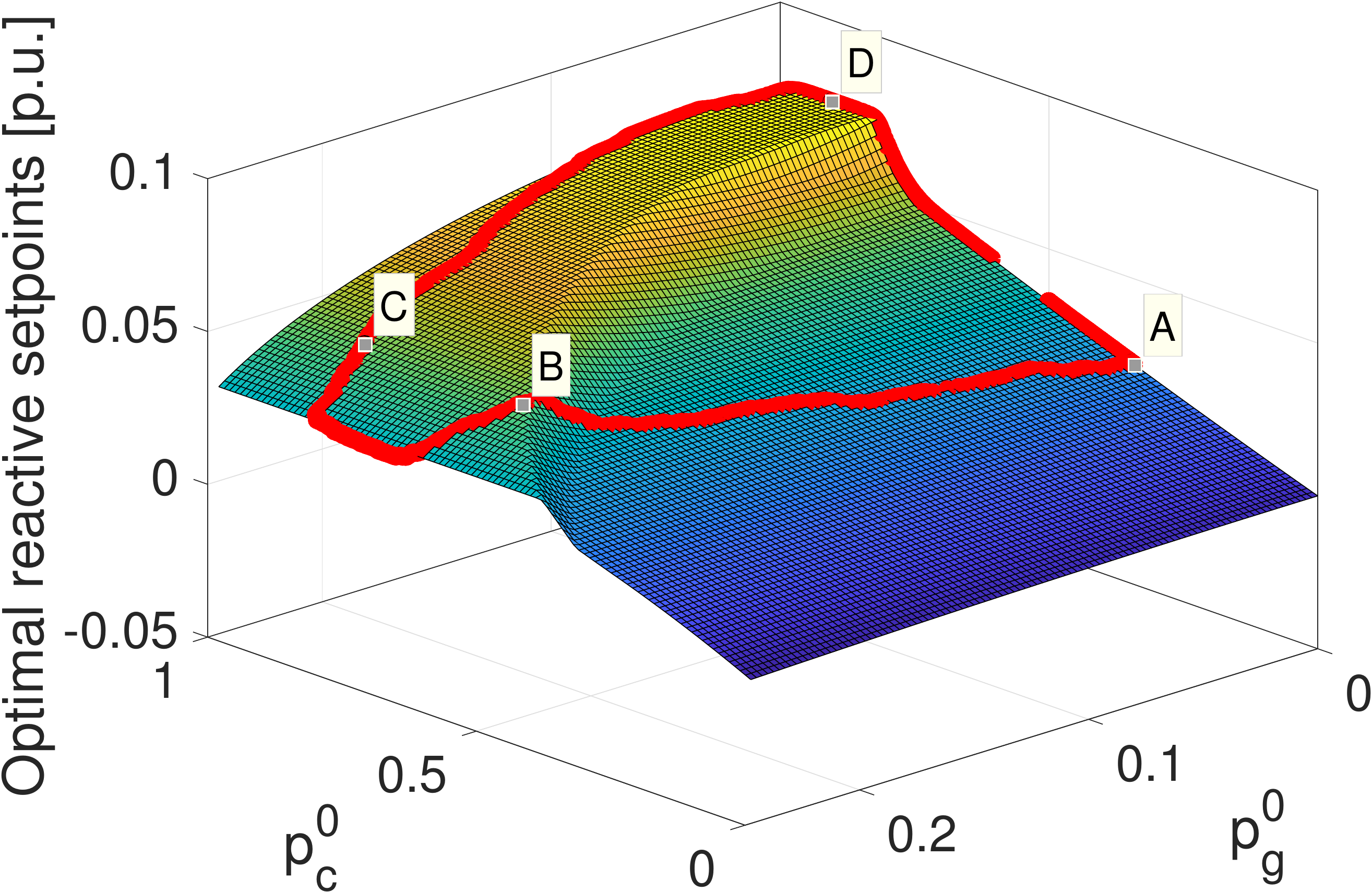}
\caption{Optimal $q_5^g$ at bus $5$ of the $13$-bus benchmark as a function of the aggregate demand $p_c^0$ and solar generation $p_g^0$. Red circles mark the optimal reactive setpoints corresponding to real-world data.}
\label{fig:duck2d}  
\end{figure}

\subsection{Two-parameter OPF on IEEE 13-bus Feeder}\label{subsec:13}
For illustrative purposes and to draw intuition, we first evaluated GP-OPF on a simple OPF setup that depended on only two parameters $(p_c^0,p_g^0)$ for the IEEE 13-bus system. We synthesized active loads and solar generation by scaling benchmark values by $p_c^0$ and $p_g^0$ respectively at all buses. Reactive loads were synthesized according to random power factors as described earlier. Thus, all OPF parameters were linear functions of $(p_c^0,p_g^0)$. To obtain the OPF mapping, we uniformly sampled $100$ values of $p_c^0$ and $p_g^0$ from $[0,1]$ and $[0,0.25]$, respectively; and then solved the related $10,000$ OPFs. Figure~\ref{fig:duck2d} illustrates the OPF mapping between $(p_c^0,p_g^0)$ and the optimal reactive setpoint for bus $5$. According to Fig.~\ref{fig:duck2d}, the optimal $q_5^g$ changes linearly with load $p_c^0$ for smaller values of $p_c^0$. This is because at low loading, voltage constraints are inactive and reactive power compensation from inverters scales roughly linearly to minimize losses. As load increases, the apparent power constraint at bus $2$ becomes active, thus causing an abrupt increase in $q_5^g$. The reactive setpoint increases until the apparent power constraint at bus $5$ becomes active, creating a nonlinear relation with solar $p_g^0$. Further, solar generation and reactive capacity are inversely related. Hence, apparent power constraints become active at higher loads for decreasing solar.

\begin{figure}[t]
\centering
\includegraphics[scale=0.22]{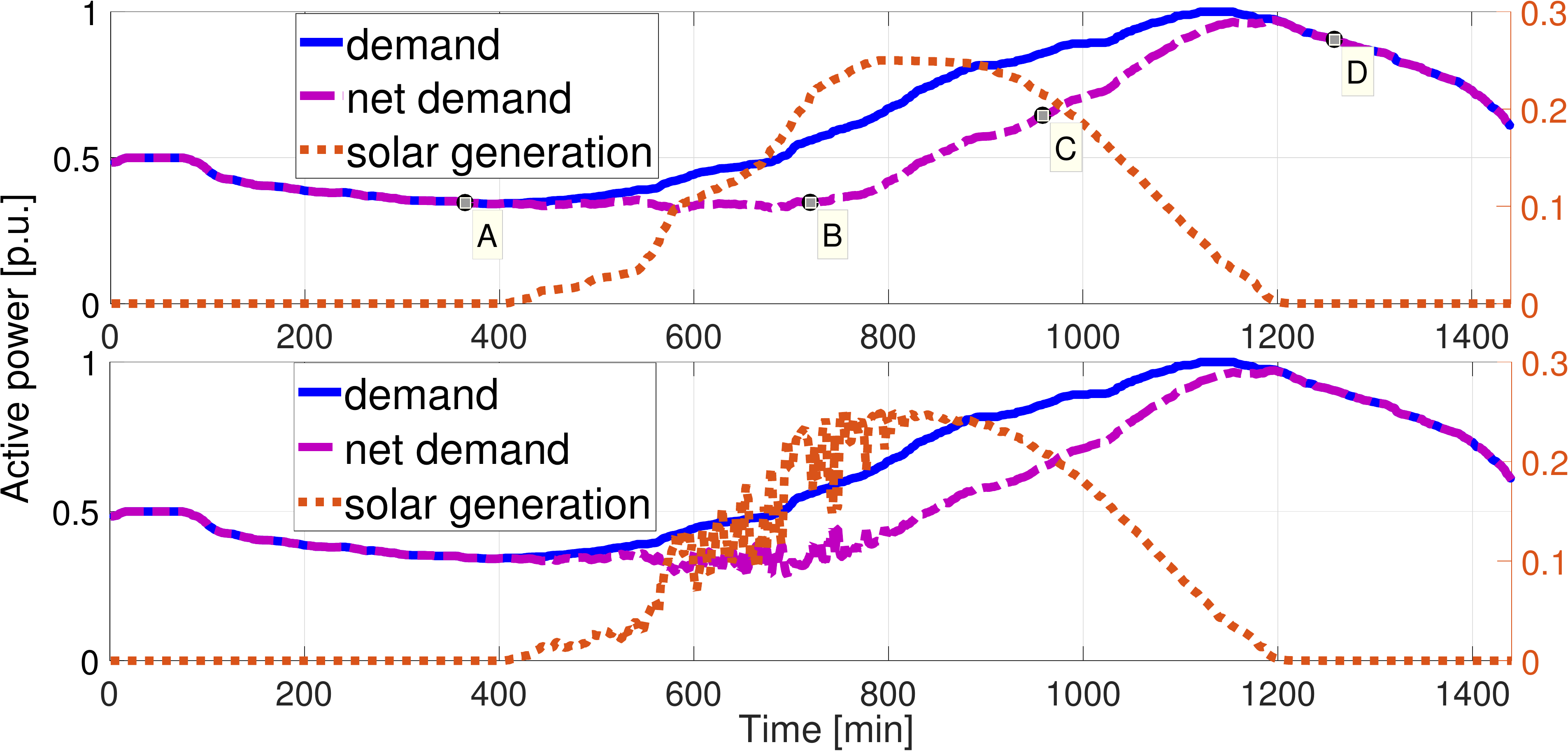}
\caption{Active demand ($p_c^0$), available active solar generation power ($p_g^0$), and net demand ($p_c^0-p_g^0$) over a day under sunny (top panel) and cloudy (bottom panel) conditions. The right axis measures the magnitude of $p_g^0$.}
\label{fig:duck_curves}  
\end{figure}

However, not all points of the OPF surface in Figure~\ref{fig:duck2d} may occur in practice. For example, as load peaks in the evening but solar during daytime, the pair $(p_c^0,p_g^0) = (1,0.25)$ is unlikely to occur. To capture more realistic grid conditions, we produced $(p_c^0,p_g^0)$ pairs using the Smart* project data by summing up loads and solar generation across all buses. The two obtained time series were smoothed using a median filter of order 150 with MATLAB's \texttt{medfilt1}, and then scaled so their peaks matched the maximum values shown in the top panel of Fig.~\ref{fig:duck_curves}. Solving the OPF for these values yielded the red dots shown in Fig.~\ref{fig:duck2d}. Points $(A,B,C,D)$ noted on Fig.~\ref{fig:duck2d} and Fig.~\ref{fig:duck_curves} (top) trace the path of optimal $q_5^g$ across time. 

\begin{figure}[t]
\centering
\includegraphics[scale=0.22]{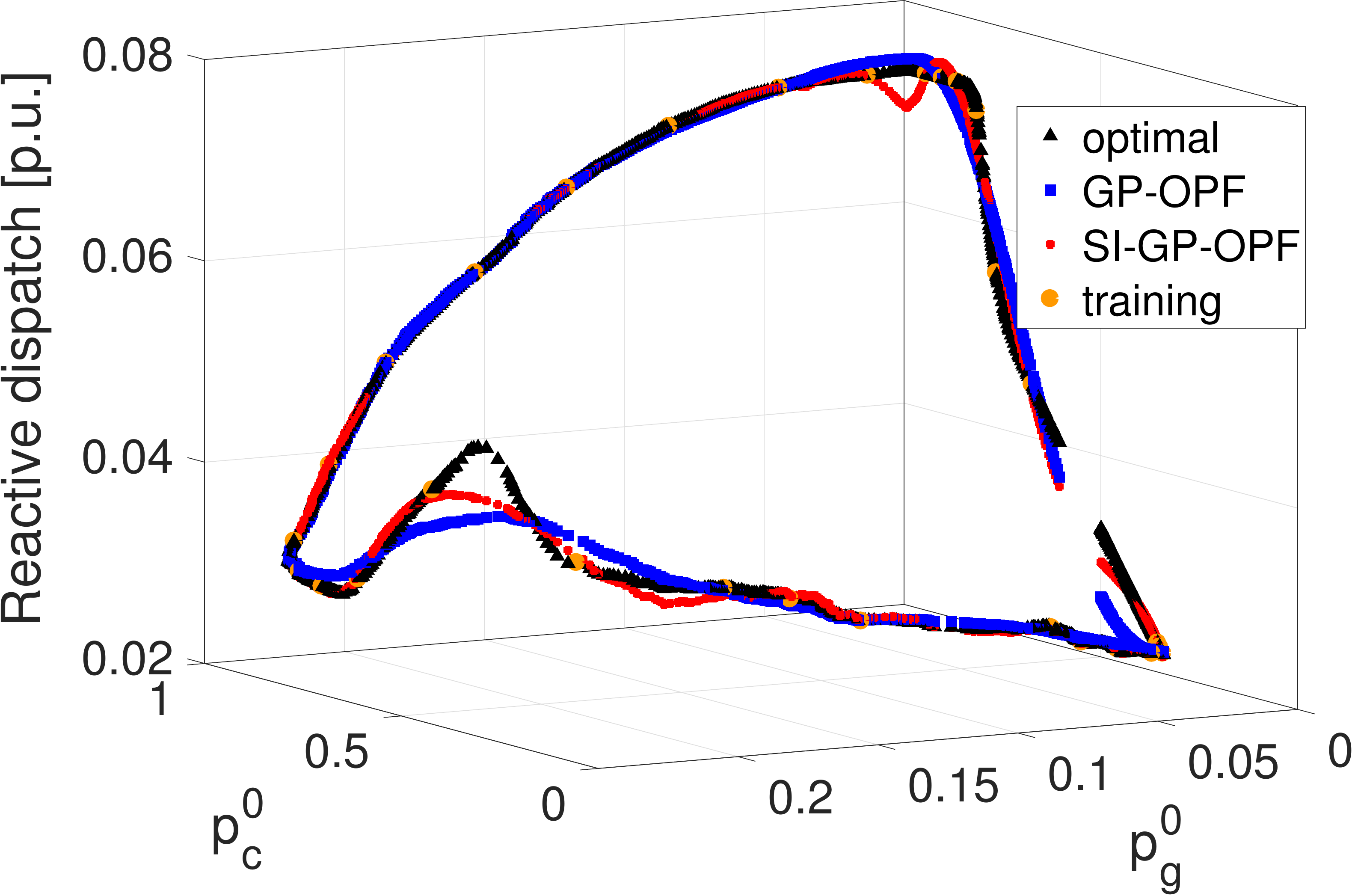}\\
\includegraphics[scale=0.22]{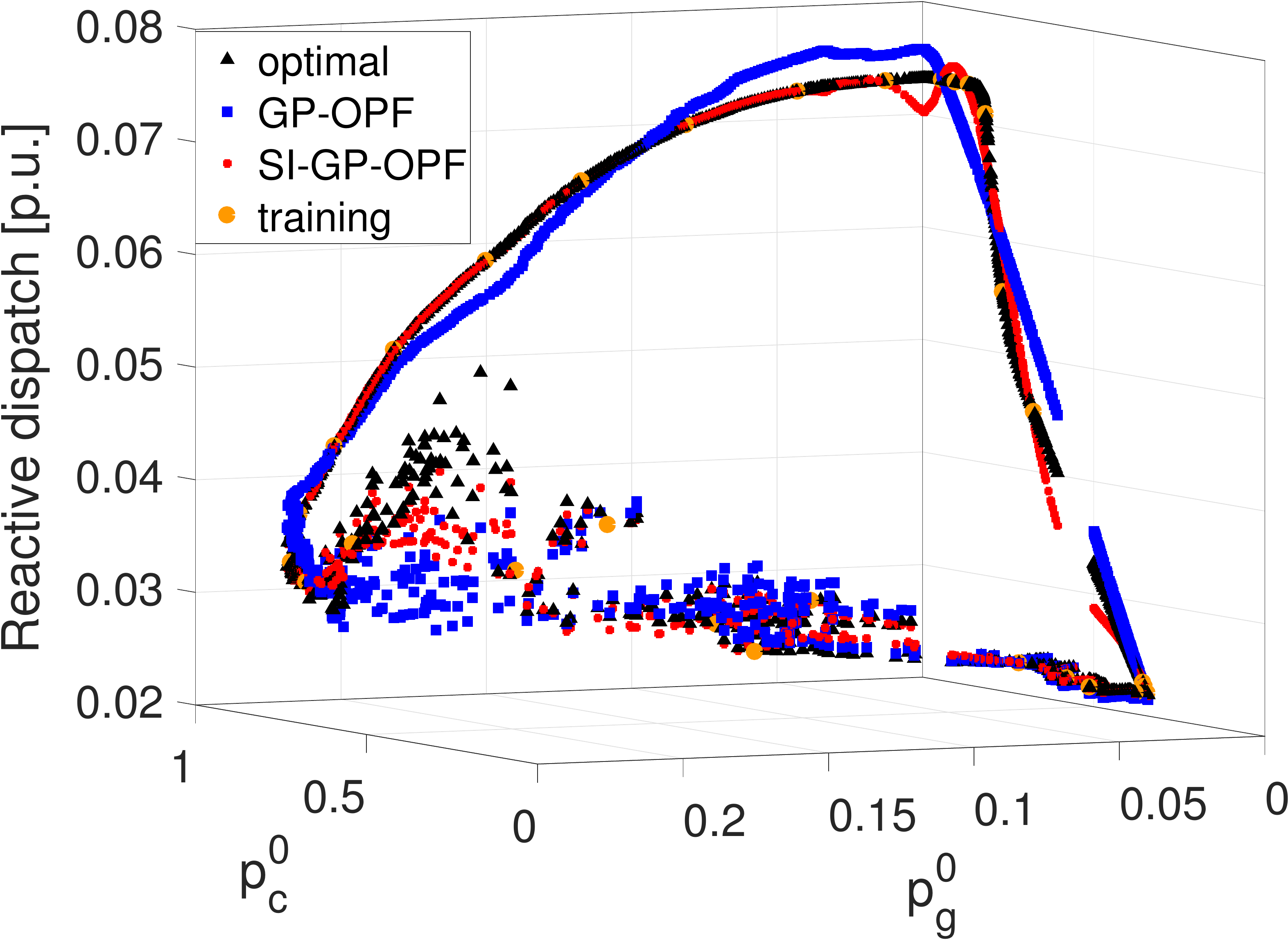}
\caption{Optimal reactive dispatch for bus $5$ of the IEEE $13$-bus benchmark and GP-OPF predictions under sunny (top) and cloudy (bottom) conditions.}
\label{fig:duck_est}  
\end{figure}

In a nutshell, it may be more practical to learn the OPF mapping across the red-dot path rather than the entire 2-D grid of Figure~\ref{fig:duck2d}. Hence, we estimated the OPF mapping along this path using GP-OPF and SI-GP-OPF. We collected $T=48$ training OPF instances by sampling the OPF path of Figure~\ref{fig:duck2d} every $40$ minutes between 6 AM till midnight. The estimated minimizers illustrated in Figure~\ref{fig:duck_est} (top panel) show that both approaches succeeded at inferring the OPF mapping. Notably, the standard deviation for the estimated $q_5^g$ across time dropped from $2.6 \times 10^{-3}$ with GP-OPF to $0.03\times 10^{-3}$ with SI-GP-OPF. The previous test simulated relatively smooth solar generation $p_g^0$. To account for solar variability, we repeated the test for the solar $p_g^0$ plotted in the bottom panel of Fig.~\ref{fig:duck_curves}. This signal was generated by summing solar generations across all buses but without smoothing them with a median filter. As seen in Figure~\ref{fig:duck_est} (bottom panel), the SI-GP-OPF outperforms GP-OPF in terms of prediction accuracy, especially in areas of larger fluctuations. In such areas, the training datapoints may not be sufficiently many for GP-OPF; as the now non-smooth solar/load trajectories were sampled every 30 minutes again. By matching the OPF mapping gradients at training points, SI-GP-OPF was able to provide more accurate predictions. 




\begin{figure}[t]
\centering
\includegraphics[scale=0.17]{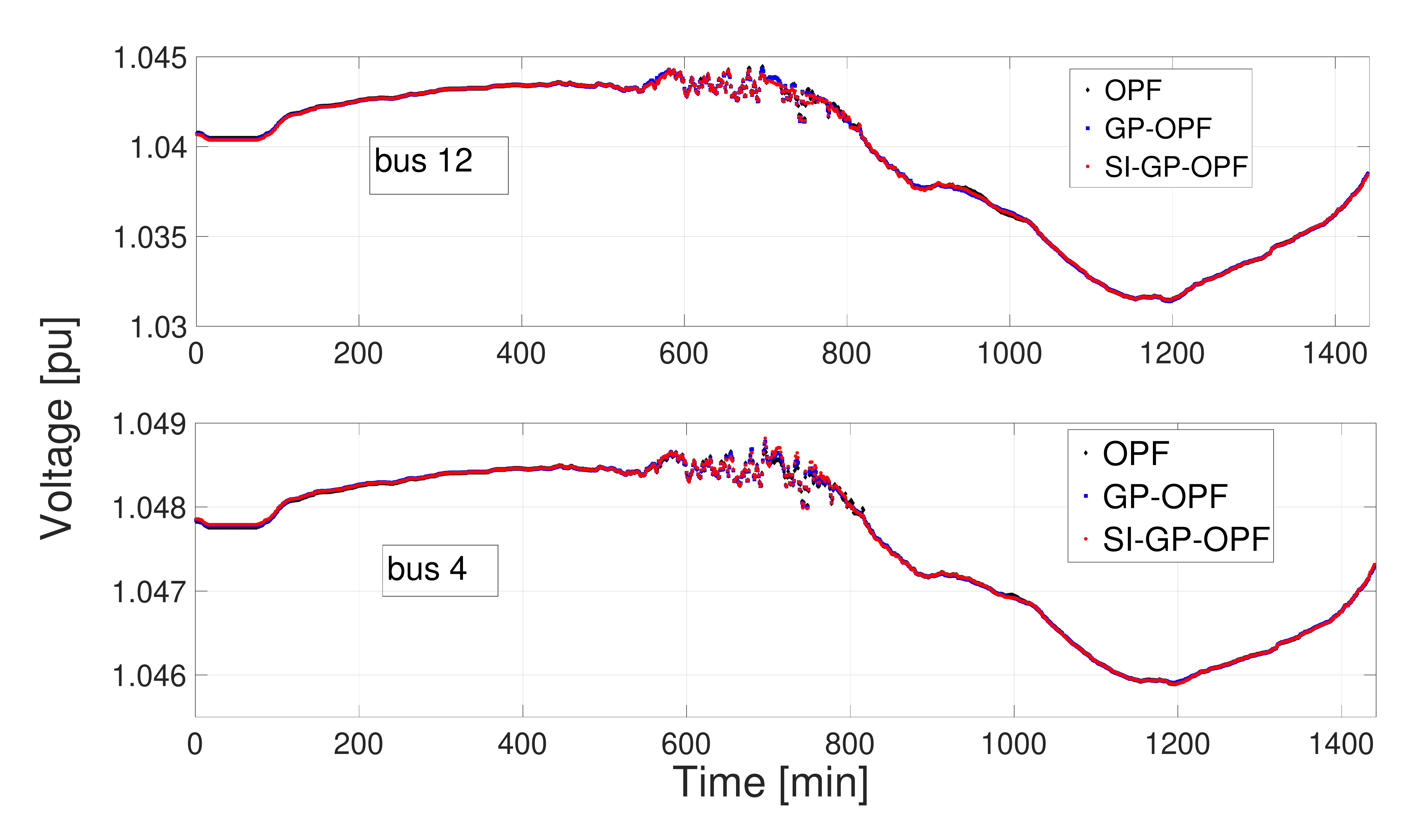}
\caption{Optimal and predicted voltages at buses 12 (top) and 4 (bottom) of the IEEE 13-bus feeder under cloudy conditions.}
\label{fig:Volt_est}  
\end{figure}

GP models can be trained for any entry of the OPF minimizer $\bx$ or other quantity of interest as long as its labels and sensitivities can be computed; they are not restricted to inverter setpoints. As an example, the previous test was repeated for inferring nodal voltages. For this test, the training data were the optimal voltages downsampled every $30$ minutes. Figure~\ref{fig:Volt_est} shows the voltage prediction results at buses $12$ (top panel) and $4$ (bottom panel) for a cloudy setup. The results confirm the effectiveness of GP-OPF and SI-GP-OPF for learning voltages across a feeder.

\subsection{Tests on IEEE 123-bus Feeder}\label{subsec:123}
GP-OPFs were also tested on the IEEE 123-bus benchmark. Here, solar and loads at each bus varied independently based on real-world data from the Smart* Project. Buses with indices that are multiplies of 5 hosted DERs yielding a total of 17 inverters. Solar generation data were normalized so their peak values matched $50\%$ of their nominal load value at that bus. Considering the period of 7:00 AM to 8:00 PM yielded 781 OPF instances related to 1-min intervals. Downsampling every $30$ minutes gave $T = 27$ training datapoints. The optimal setpoints for all 17 inverters were learned using the three methods: GP-OPF, SI-GP-OPF, and RF-SI-GP-OPF. For the last one, the number of RFs was set to $D=1,600$ striking a good trade-off between learning time and accuracy.

\begin{figure*}[t]
	\centering	
	{\includegraphics[width=0.22\textwidth]{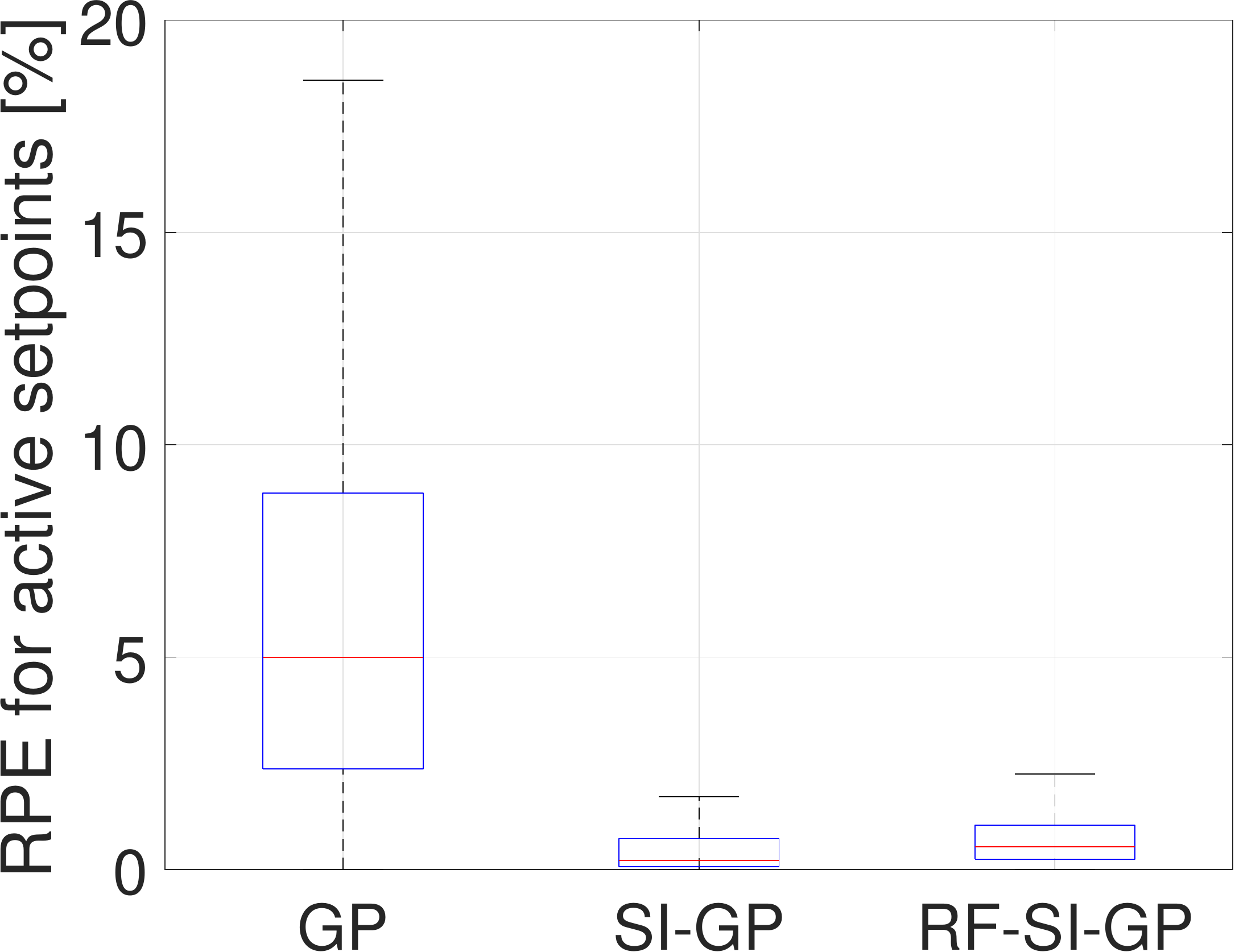}}
	\hspace*{+1em}
	{\includegraphics[width=0.22\textwidth]{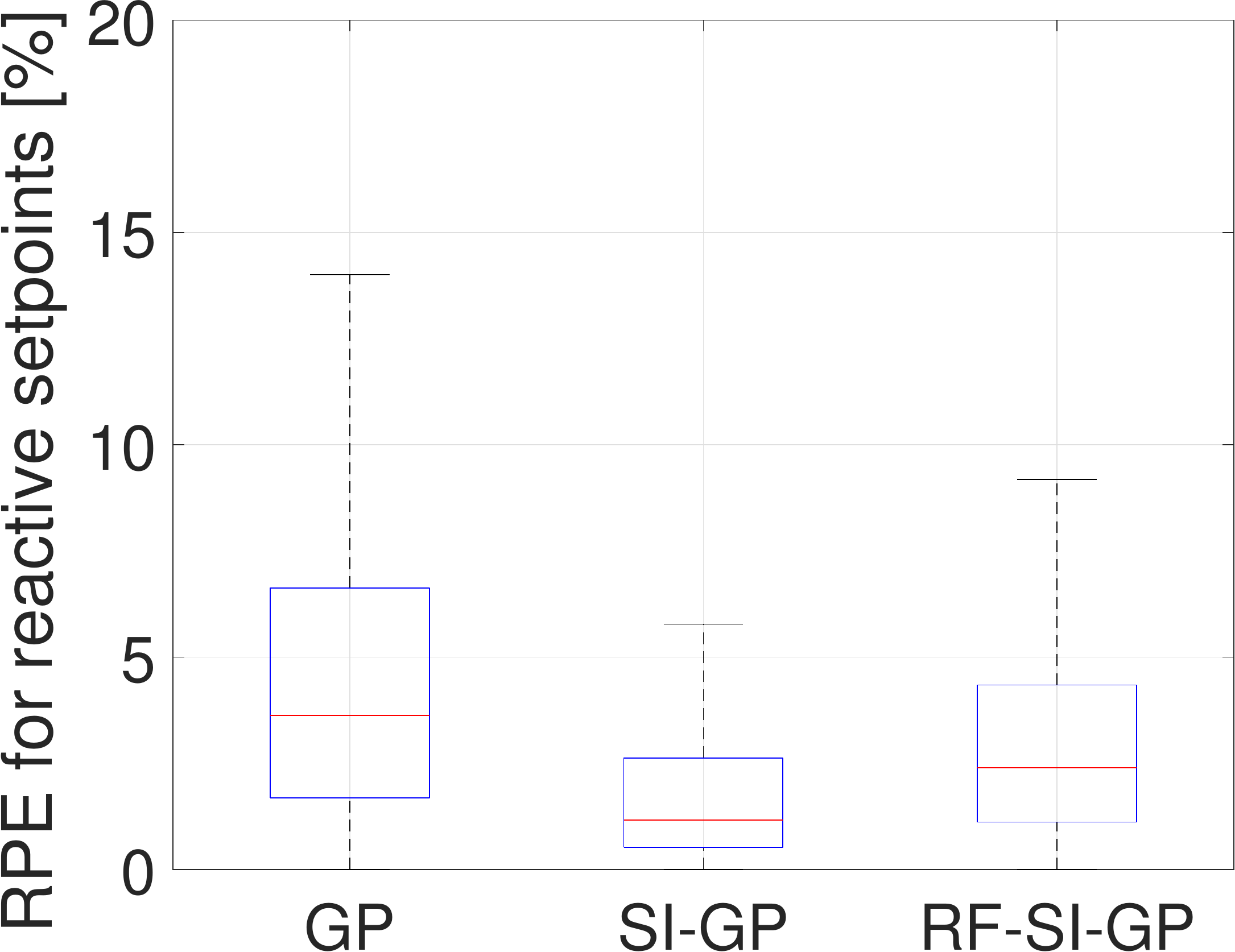}}
	\hspace*{+1em}
	{\includegraphics[width=0.22\textwidth]{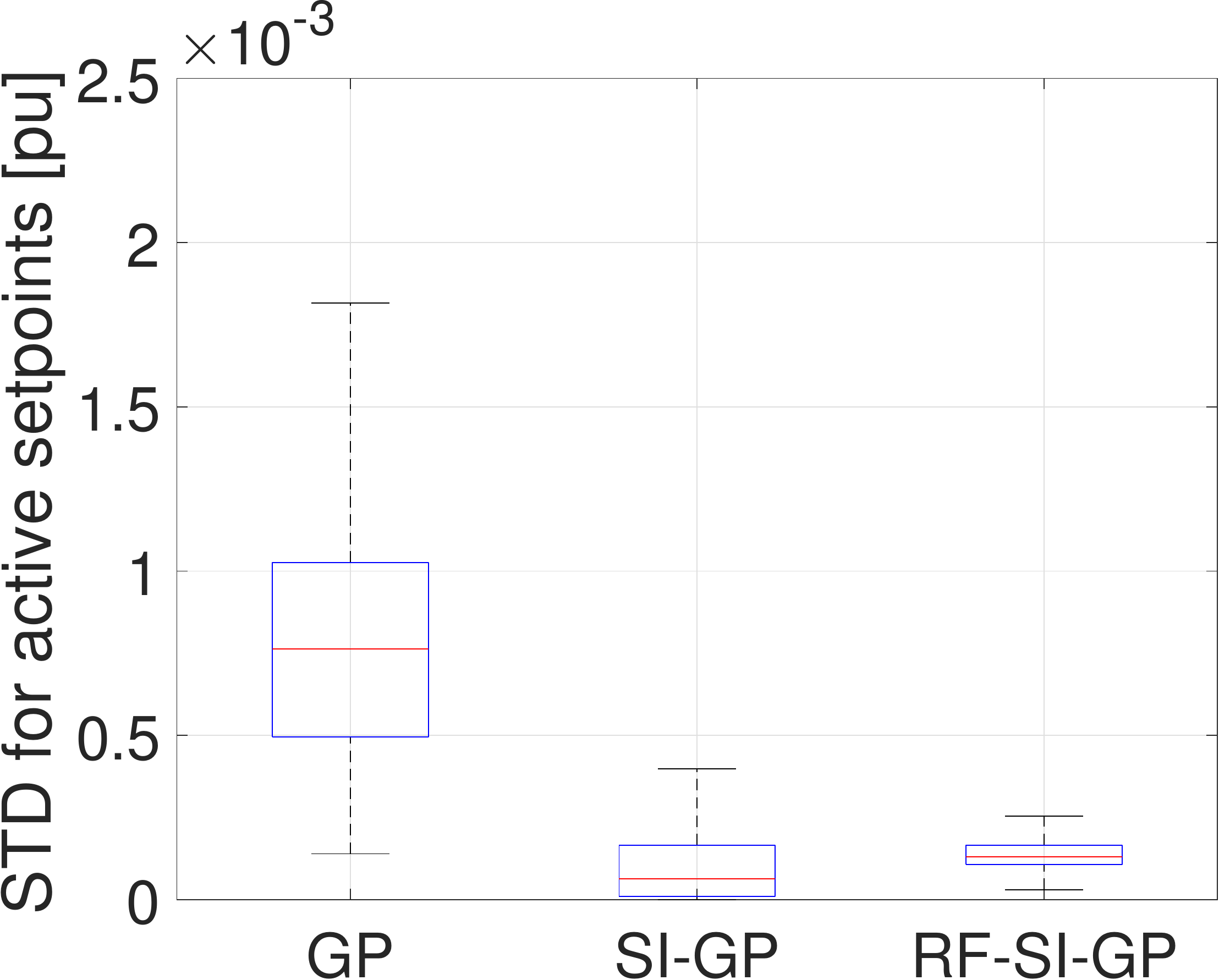}}
	\hspace*{+1em}
	{\includegraphics[width=0.22\textwidth]{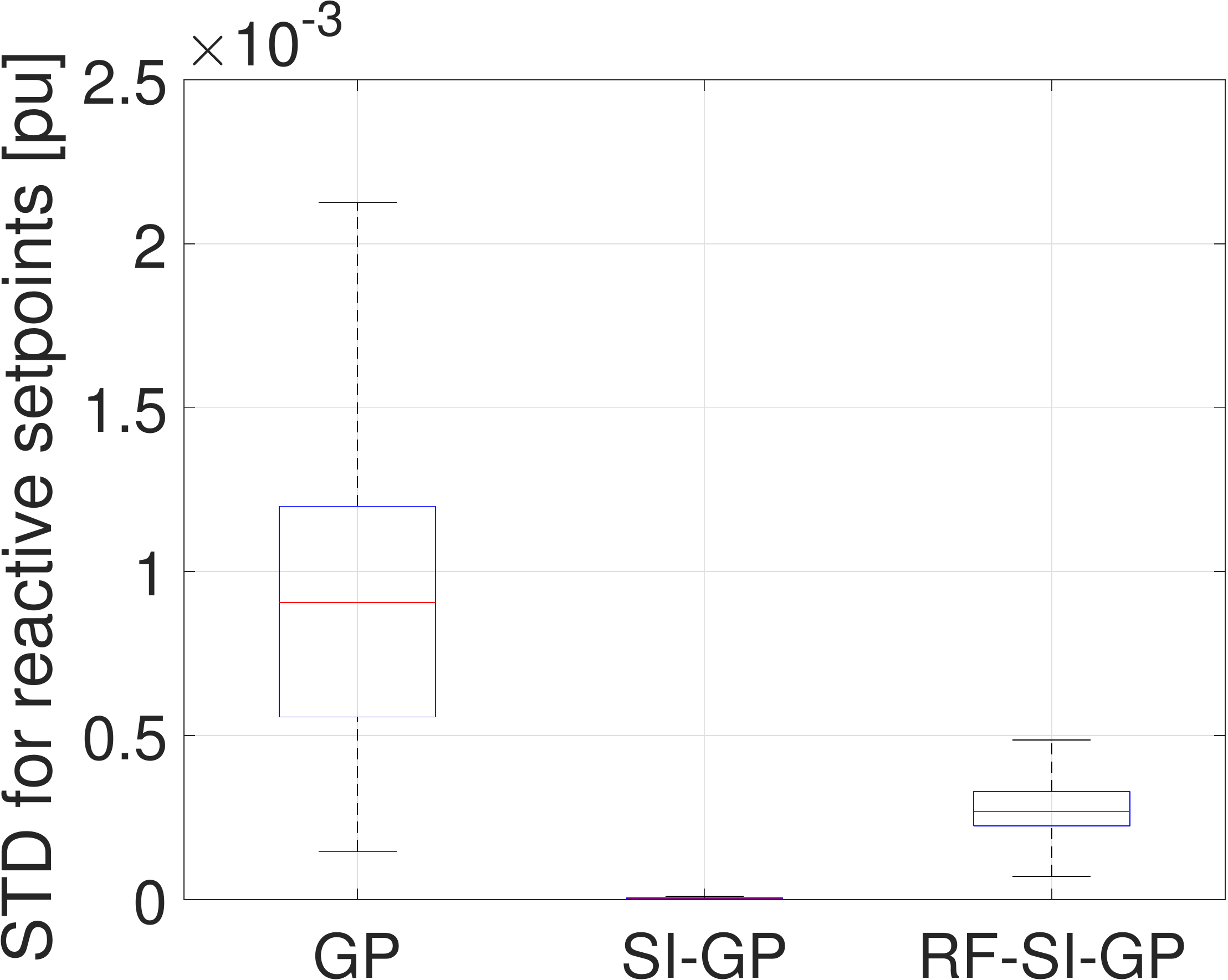}}
	\caption{Boxplots for RPEs and standard deviations (STD) while estimating optimal (re)active setpoints.}
	\label{fig:boxplots}
\end{figure*}

Figure~\ref{fig:boxplots} shows the boxplots for RPEs of optimal (re)active setpoints for all inverters and time instances. It also shows their standard deviations (STDs) as predicted by the GP models. Overall, SI-GP-OPF outperforms GP-OPF in terms of error and uncertainty, which confirms the advantage of incorporating sensitivities into learning. SI-GP-OPF is significantly better than GP-OPF primarily for active power setpoints. It is also evident that the RF-based approximation entails some performance degradation in exchange of computational speed-up to be detailed later. 

\begin{figure}[t]
	\centering
	\includegraphics[scale=0.22]{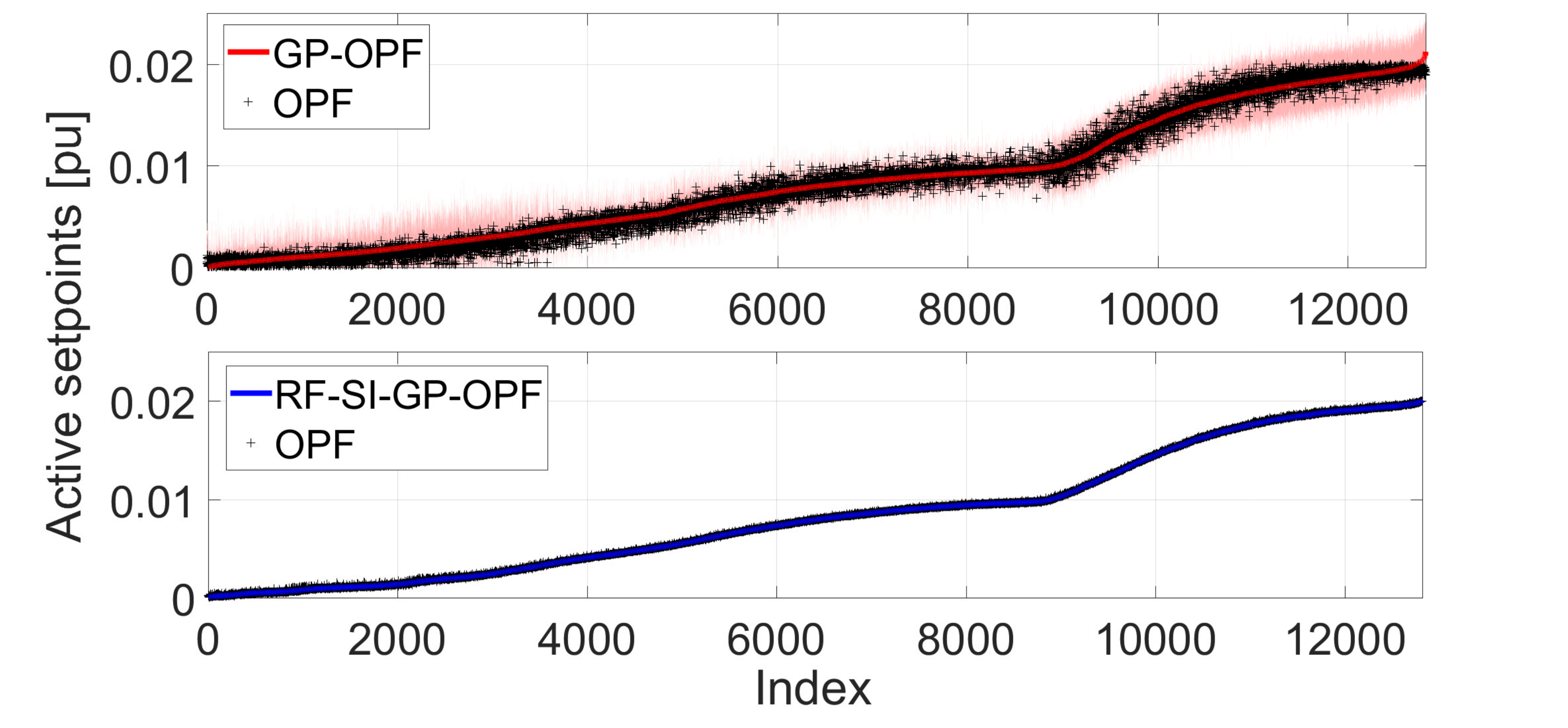}
	\includegraphics[scale=0.22]{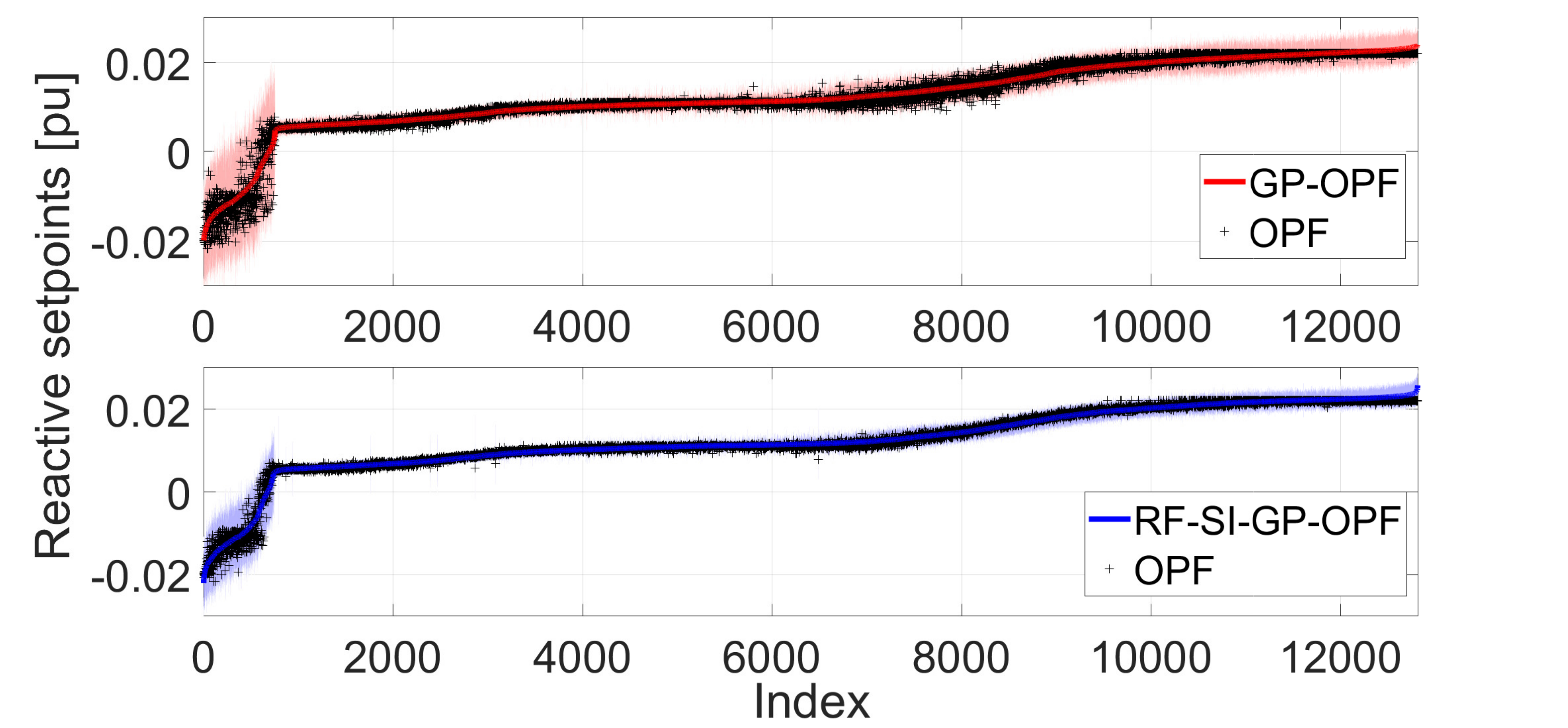}
	\caption{{Sorted optimal and estimated active (top) and reactive (bottom) setpoints over inverters and instances using GP-OPF and RF-SI-GP-OPF.}}
	\label{fig:pg_All}  
\end{figure}

\begin{figure}[t]
	\centering
	\includegraphics[scale=0.22]{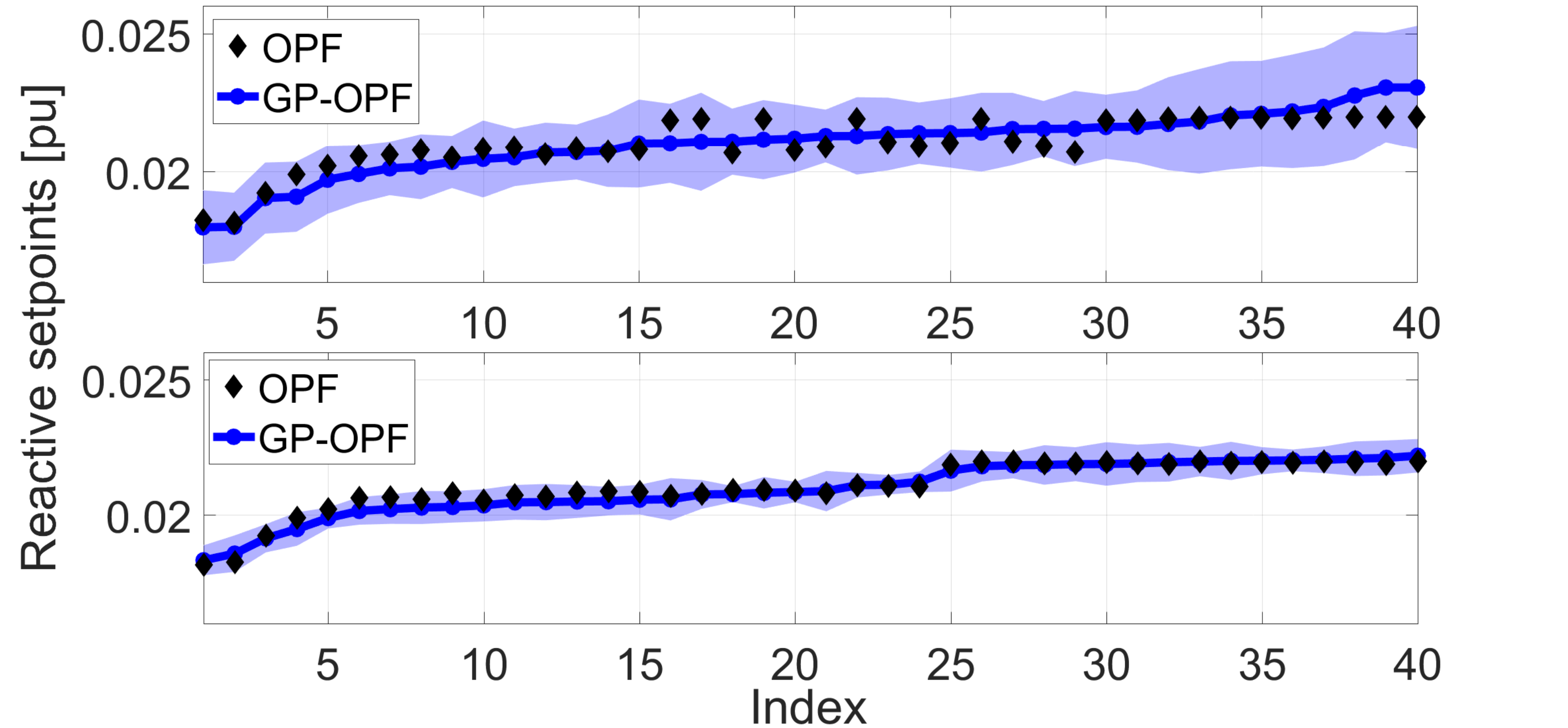}
	\caption{{Optimal and predicted reactive setpoints for inverter 10 over testing instances using GP-OPF. Instances have been sorted in increasing order based on predictions for presentation purposes. Shaded area corresponds to uncertainty interval of $\pm2\sigma$ computed from~\eqref{eq:mmse:c}.}}
	\label{fig:cluster}  
\end{figure}

While RPEs provide relative information on the estimation accuracy, insight is needed on how closely the estimated values follow the optimal setpoints. Figure~\ref{fig:pg_All} depicts the optimal setpoints and their predictions across all inverters and instances of the testing dataset. For clarity of presentation, the points have been sorted in increasing order with respect to the {predicted} value. The learned setpoints were obtained using RF-SI-GP-OPF. The shaded areas demonstrate the {$\pm 3\sigma$} confidence interval obtained by taking the square root of the diagonal entries of the covariance matrix in~\eqref{eq:mmse:c}. These results visualize the improvement achieved by using sensitivities. Figure~\ref{fig:pg_All} shows high uncertainty intervals and low accuracy near negative reactive power setpoints. This is due to insufficient training data for over-voltage conditions. This can be solved by adding more labeled data in such instances. Figure~\ref{fig:cluster} will later show how adding pertinent training data can solve such issues.

\begin{figure}[t]
	\centering
	\includegraphics[scale=0.24]{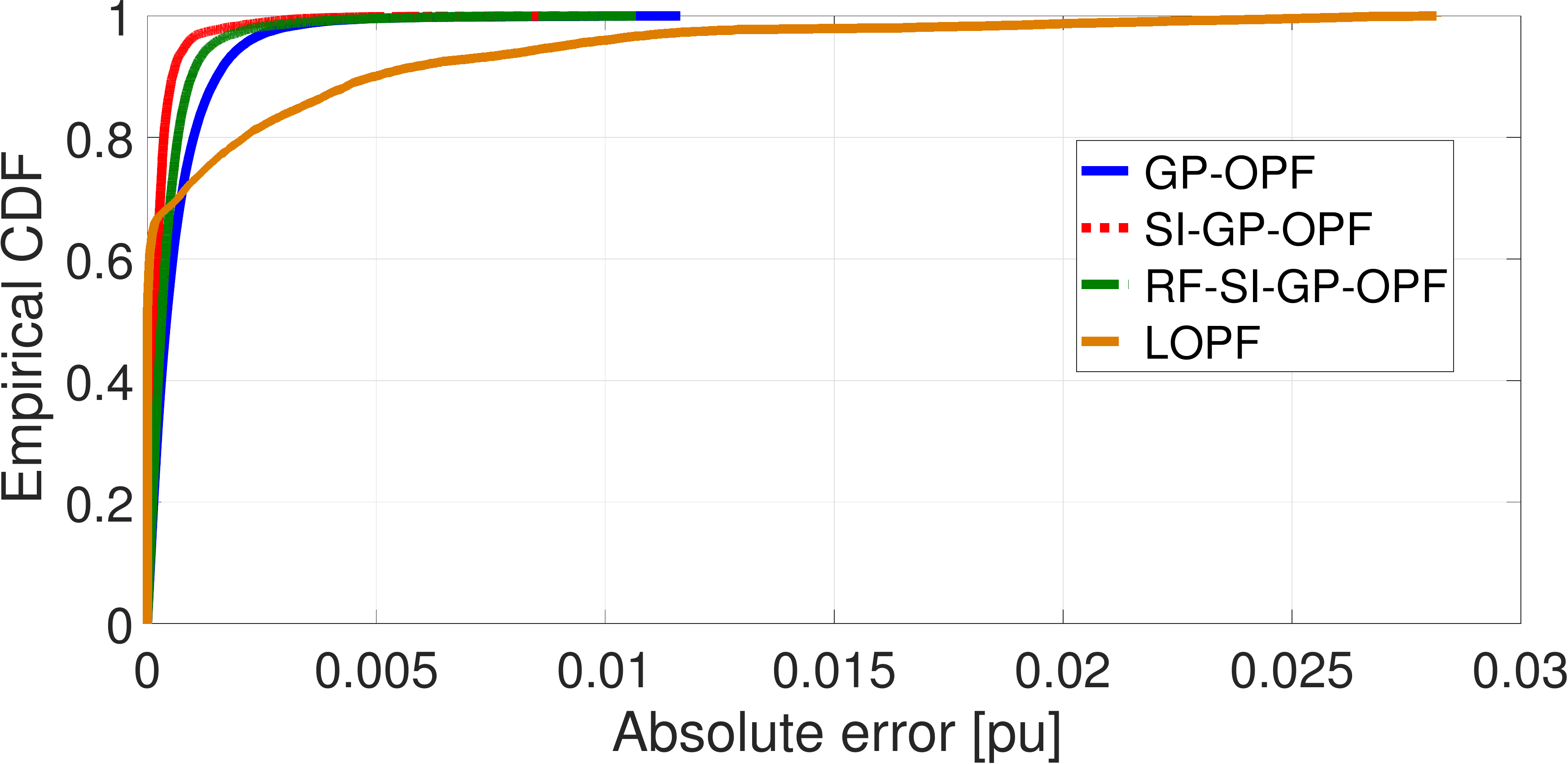}
	\caption{{Empirical CDF of error for predicting reactive setpoints.}}
	\label{fig:cdf}  
\end{figure}

One may argue that nearly optimal inverter setpoints can be obtained by a linearized OPF, which enjoys improved computational complexity over the exact SOCP formulation of the OPF. To this end, we compared GP-OPF with the linearized OPF (LOPF) detailed in Appendix~\ref{sec:LDF}. The LOPF in~\eqref{eq:LDF} approximates the AC-OPF of~\eqref{eq:OPF} by a quadratic program. LOPF does not model line currents explicitly, and hence, the current limits of~\eqref{eq:OPF:lmax} were dropped. This does not harm the comparison since line limits of~\eqref{eq:OPF:lmax} were not binding in our tests. We solved LOPF under the same grid conditions as in the previous tests. The average running time per LOPF instance was $0.292$~seconds, which is over $20$ times the prediction time of RF-SI-GP-OPF. Figure~\ref{fig:cdf} shows the empirical cumulative distribution function (CDF) of the absolute error of reactive setpoints obtained from LOPF, GP-OPF, SI-GP-OPF, and RF-SI-GP-OPF. The GP-based approaches were trained using $T=27$. Figure~\ref{fig:cdf} confirms the superior accuracy of a well-trained GP-OPF model over LOPF. This behavior was expected because the GP-based schemes are trained to follow the optimal setpoints, whereas the LOPF is solving an approximate optimization problem.

\begin{figure}[t]
	\centering
	\includegraphics[scale=0.14]{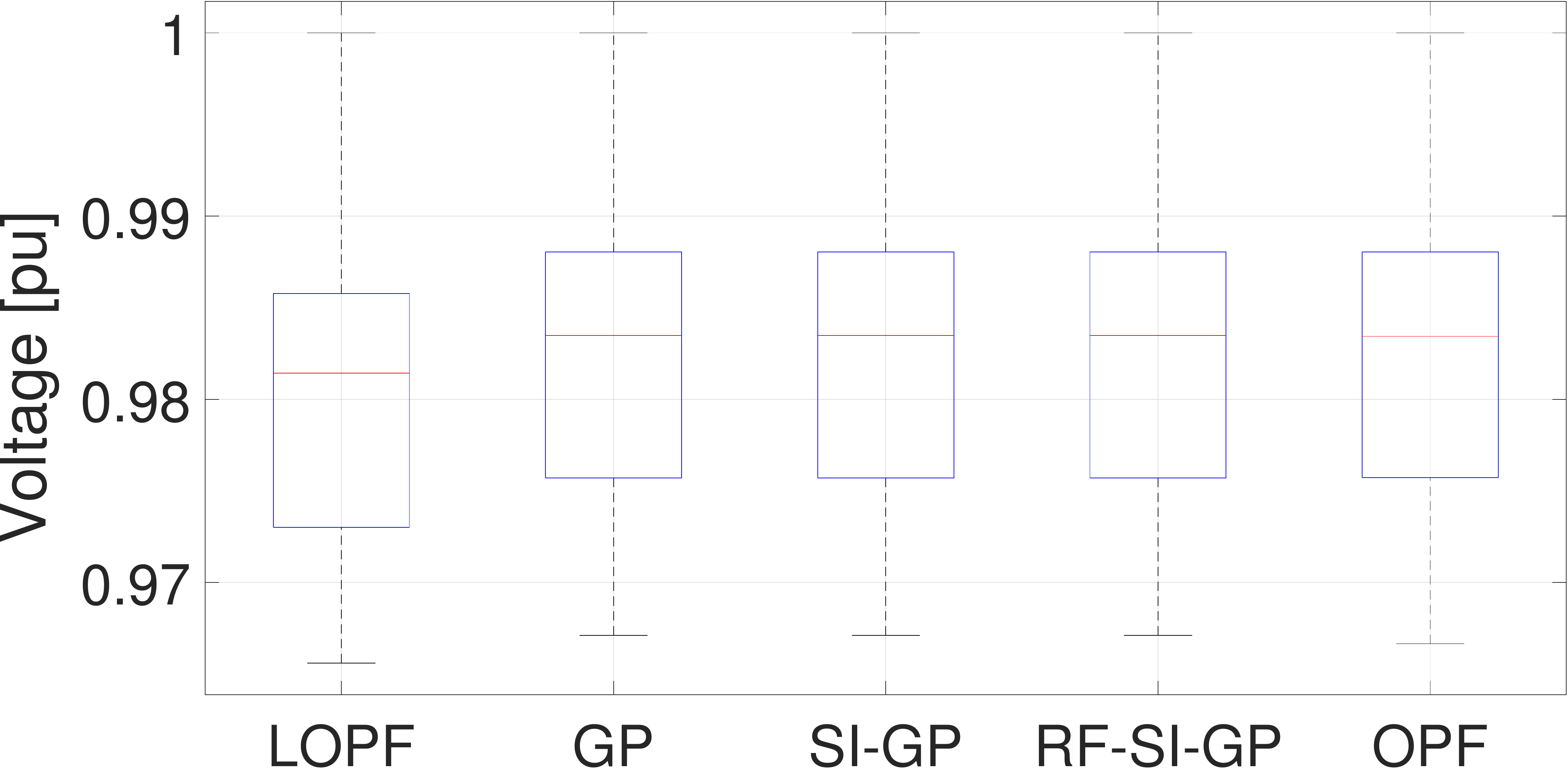}
	\caption{{Voltage across buses and testing instances obtained from solving the PF problem using the inverter dispatches obtained using different approaches.}}
	\label{fig:PF}  
\end{figure}

To verify feasibility of the setpoints predicted by GP-OPF and the setpoints computed by LOPF, we plugged these setpoints into the AC power flow equations and computed the induced grid voltages. To ease comparison, the substation voltage was set to $v_0=1$~pu. Figure~\ref{fig:PF} shows boxplots of voltages across buses and testing instances. The results confirm that the error of the learned setpoints propagated through PF equations, does not render any infeasibility. Further, the GP-based methods yield lower voltage deviations than the LOPF.

\begin{figure}[t]
	\centering
	\includegraphics[scale=0.22]{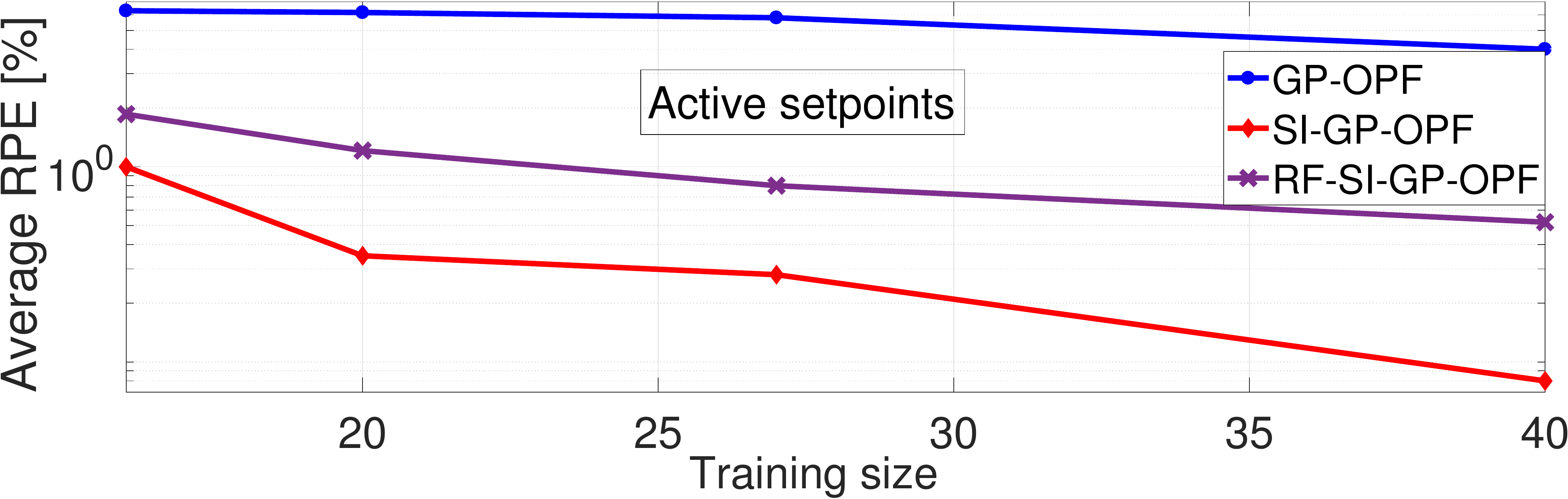}
	\includegraphics[scale=0.22]{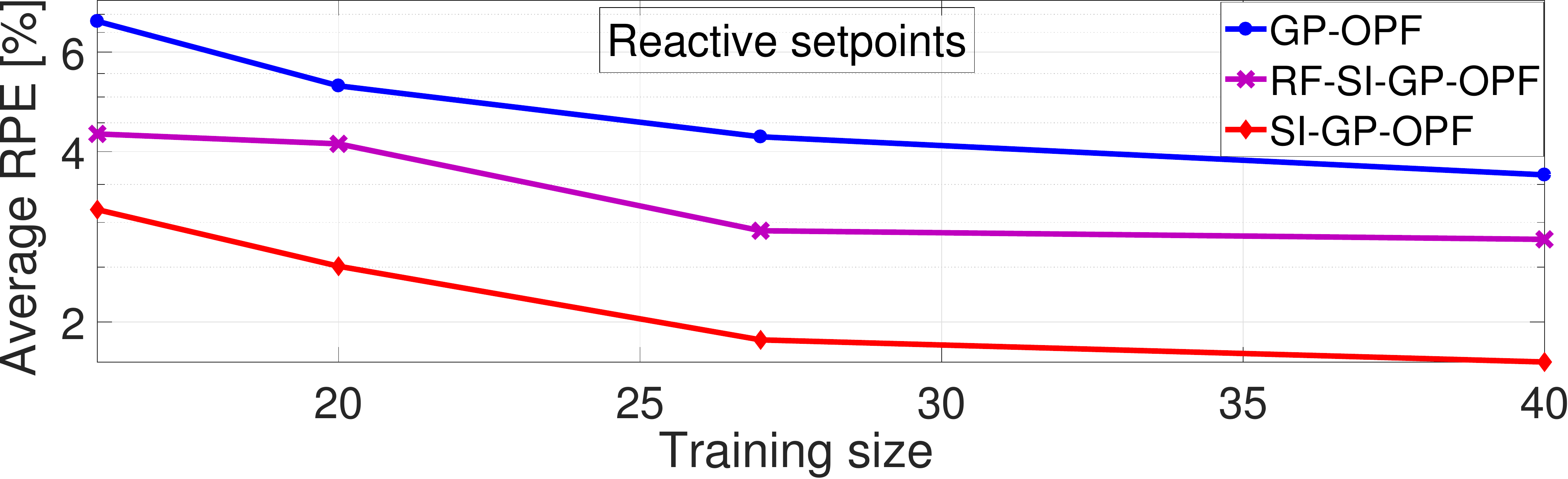}
	\caption{Average RPE while estimating active (top) and reactive setpoints (bottom) using training datasets of different size $T$.}
	\label{fig:RPE_T}  
\end{figure}

\begin{figure}[t]
	\centering
	\includegraphics[scale=0.21]{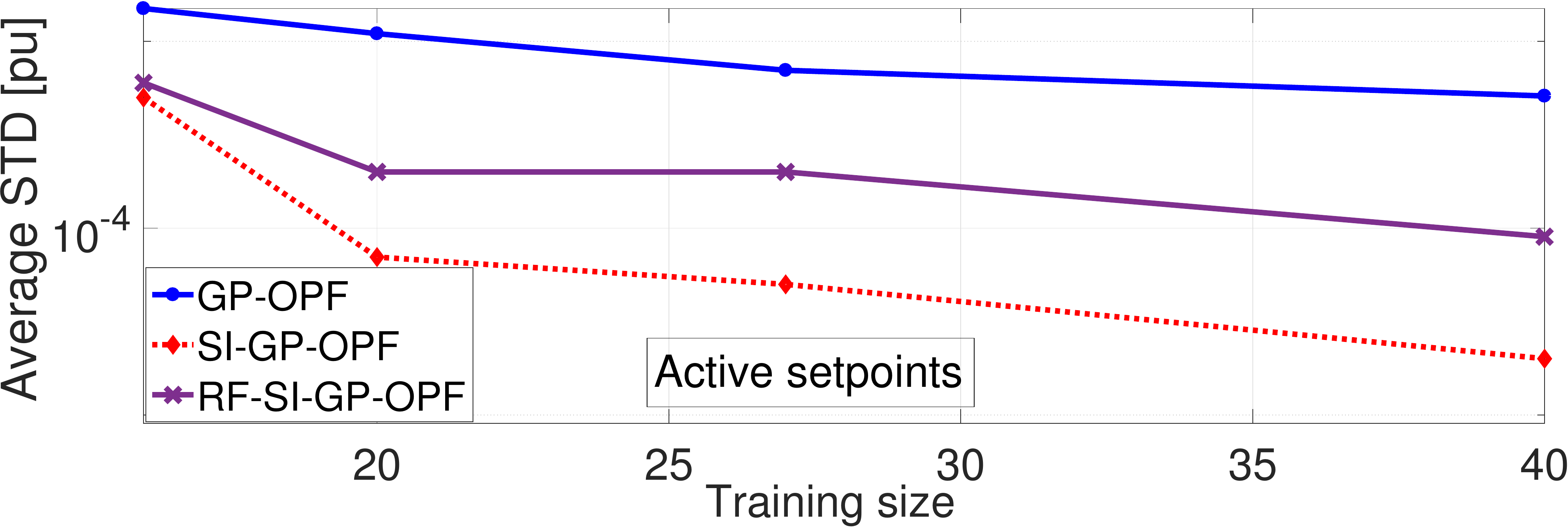}
	\includegraphics[scale=0.21]{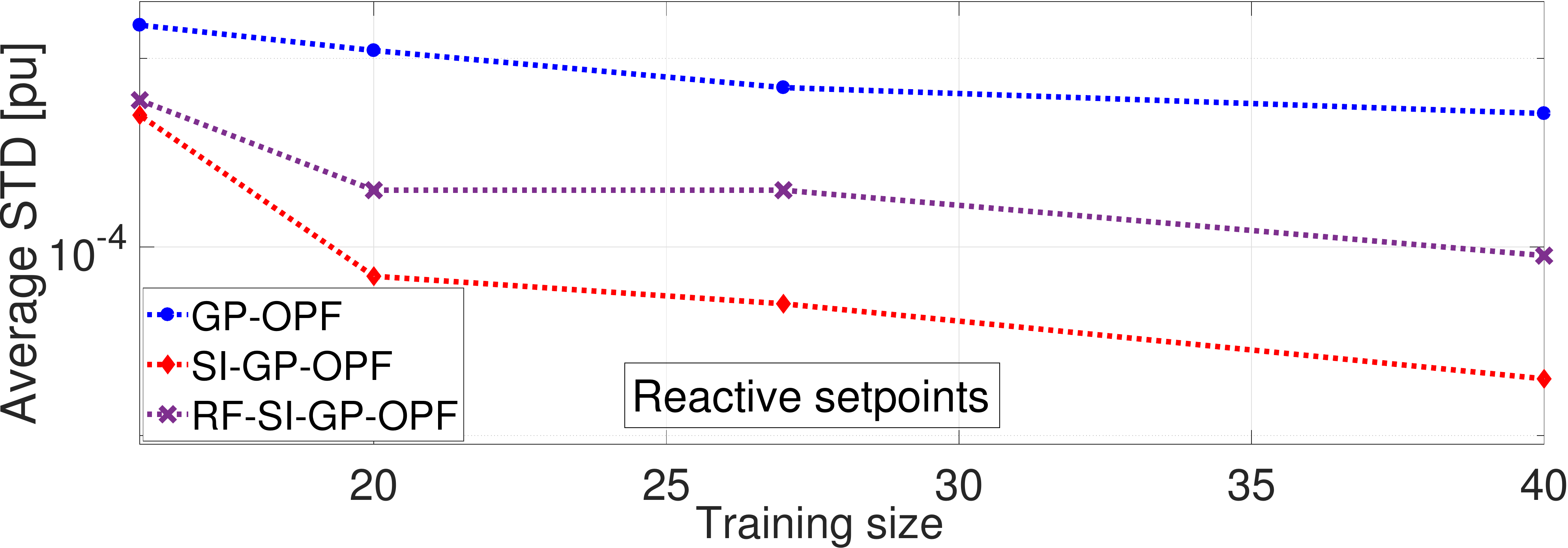}
	\caption{Average standard deviation (STD) while estimating active (top) and reactive setpoints (bottom) using training datasets of different size $T$.}
	\label{fig:sigma_T}  
\end{figure}

To study the effect of training size $T$, the inverter dispatches were learned for $T \in \{16,20,40\}$ by uniformly sampling the labeled data every $\{50,40,20\}$ minutes, respectively. To compare the estimation performance over $T$, we averaged RPEs and STDs over inverters and instances. Figures~\ref{fig:RPE_T} and~\ref{fig:sigma_T} show the results for predicting active (top) and reactive (bottom) setpoints. Table~\ref{tbl:pred_times} reports the average time of predicting setpoints across the feeder per OPF instance for each $T$. These times of course do not account for data generation. All learning methods are faster than solving an OPF, which takes $3.8$~seconds per instance. The computation time for finding the sensitivities was $0.07$~seconds, which is negligible compared to the OPF time. The tests further confirm that prediction accuracy improves with increasing $T$ for all methods. Moreover, the sensitivity-informed GP-OPF featured lower RPE/STD than plain GP-OPF across $T$. For reactive setpoints, the RPE attained by GP-OPF using $T=40$ is achieved by RF-SI-GP-OPF using less than $T=25$.

\begin{table}[t]\label{tbl:pred_times}
	\renewcommand{\arraystretch}{1.2}
	\caption{Average running times [s] for predicting optimal setpoints for all inverters of the IEEE 123-bus system per OPF instance.}
	\centering
	\begin{tabular}{|l|r|r|r|r|}
		\hline\hline
	    \multirow{2}{*}{  } & \multicolumn{4}{c|}{Size of Training Dataset $T$}   \\
	   \cline{2-5}
	    & $40$ & $27$ & $20$ & $16$\\
		\hline\hline {GP-OPF} & $0.004$ & $0.002$ & $0.002$ & $0.001$\\
		\hline	{SI-GP-OPF}& $0.853$ & $0.388$ & $0.228$ & $0.141$ \\
		\hline	{RF-SI-GP-OPF}&  $0.012$ & $0.012$& $0.011$  & $0.011$ \\
		\hline\hline
	\end{tabular}
\end{table}

\begin{figure}[t]
	\centering
	\includegraphics[scale=0.22]{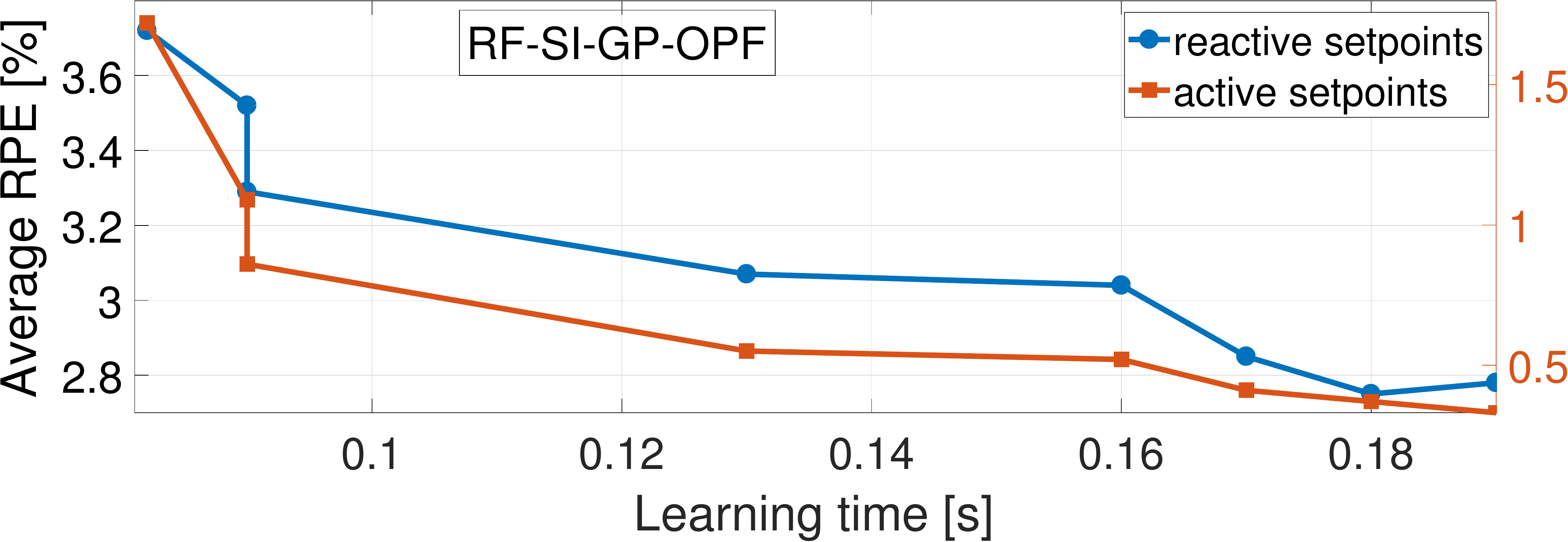}
	\caption{Average RPE over prediction time for estimating active (right axis) and reactive (left axis) setpoints using RF-SI-GP-OPF. The number of random features $D$ was increased (left to right) across the range $[600,2000]$ with increments of $200$. Larger $D$ yield smaller errors but longer times.}
	\label{fig:D_RPE}  
\end{figure}

Fixing $T=27$, we studied the effect of $D$ on running time and accuracy. The number of RFs varied across $[600,2000]$ with increments of $200$. Figure~\ref{fig:D_RPE} shows the average RPE for inferring setpoints using RF-SI-GP-OPF. As anticipated, prediction time increases with $D$, whereas RPE is decreasing with $D$. Increasing $D$ from $1,600$ to $2,000$ offers marginal prediction improvement. These results show the trade-off between accuracy and prediction time using RFs.  

We subsequently explored the practical merit of uncertainty intervals. High predictive variances indicate that particular grid conditions $\btheta$'s were not well represented in the training set. Therefore, GP-OPF predictions with large variances identify areas in the parameter space $\btheta$ from which more labels $y(\btheta)$ need to be sampled. To validate our hypothesis, we used K-means to cluster $\btheta$'s into 20 clusters. We trained a GP-OPF model using training data drawn only from 15 out of the 20 clusters. Testing data were selected by choosing 8 samples per each of the remaining 5 clusters. This sampling protocol ensured that testing data differed from training data. Figure~\ref{fig:cluster} (top) depicts optimal and predicted reactive setpoints for inverter 10. For better presentation, setpoints have been sorted based on the predicted value. We repeated the test but now included training data from all 20 clusters. Figure~\ref{fig:cluster} (bottom) shows the reactive setpoints for the same inverter and over the same OPF instances as those on the top panel. Of course, these testing instances were not part of the training dataset. As expected, the predictions of the bottom panel are closer to the optimal setpoints, and that is also verified by their smaller predictive variances.

\begin{figure}[t]
\centering
\includegraphics[scale=0.15]{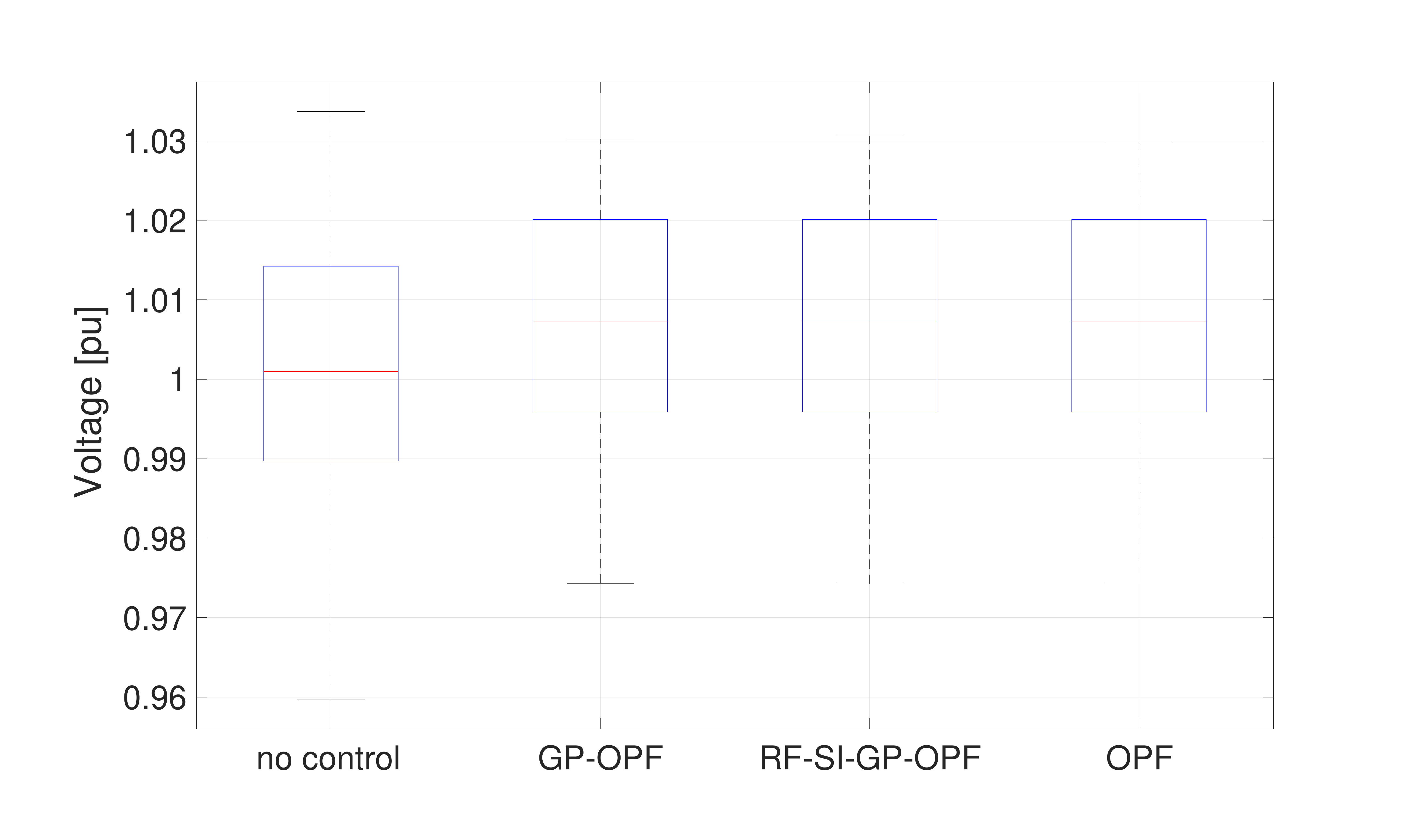}
\caption{Boxplots of AC voltages obtained under no inverter control, GP-OPF, RF-SI-GP-OPF, and AC-OPF.}
\label{fig:NC_V}  
\end{figure}

To analyze the performance of the proposed GP-based schemes under more extreme grid conditions, the solar active power generation was further increased as follows: \emph{a)} Inverters were placed at all non-zero-injection buses, thus increasing the total number of inverters to $85$; and \emph{b)} Load/solar injections were scaled to match the nominal benchmark values and twice the benchmark values, respectively. Under this setting, voltage deviations exceeded the $\pm 3\%$ per unit (pu) limit. Note that the ANSI-C.84.1 standard specifies a $\pm 5\%$ pu deviation limit for service voltages, i.e., voltages at the customer connection point. To account for the voltage drop between the service and distribution transformers, we adopted the common practice of aiming for a maximum of $\pm 3 \%$ voltage deviation at distribution transformers; see e.g.,~\cite{kersting}. For this test, we generated more data by synthetically generating a set for a period of two days rather than one. This was accomplished by perturbing the data described in the first paragraph of Section~\ref{sec:tests} with additive random Gaussian noise with variance of $0.001$ pu. The training dataset was obtained by downsampling every $5$ minutes of both days of data. The remaining instances from the second day were chosen as the testing dataset. The GPs were modeled using a Gaussian covariance with automatic relevance determination (ARD). The GP-OPF and RF-SI-GP-OPF yielded inverter setpoints that were subsequently used to solve the nonlinear PF equations. Figure~\ref{fig:NC_V} compares the obtained voltages with the voltages induced by an AC-OPF solution as well as with the voltages under no inverter control. The plots demonstrate that without inverter control, voltages exceed the $\pm 3\%$ deviation limits, whereas the GP-based schemes cause acceptable voltage deviations. 

\color{black}

\section{Conclusions}\label{sec:conc}
The GP-OPF is a fast and effective method for learning the OPF mapping for expedited inverter control. For large datasets, RF-GP-OPF features reduced computational complexity over GP-OPF. SI-GP-OPF improves estimation accuracy at the expense of increasing the training and operation speed, yet its RF-based counterpart reduces both training and prediction time. If a labeled training dataset of OPF solutions is available, GP-OPF predicts inverter setpoints faster than RF-SI-GP-OPF, though RF-SI-GP-OPF yields more accurate results. If a dataset is not available, then RF-SI-GP-OPF outperforms GP-OPF both in prediction accuracy and computational time. Both learning methods are considerably faster than solving the OPF.

Extensive numerical tests on the IEEE 13- and 123-bus feeders corroborate our findings. In particular, GP-OPF predicted near-optimal setpoints of the 123-bus system within $0.005$ seconds, while solving the OPF took $3.8$ seconds. SI-GP-OPF achieved the same prediction accuracy as GP-OPF by using only $1/4$ of the OPF instances (training labels). Computing the required sensitivities was posed as solving a set of linear equations, which took $0.07$ seconds. Finally, RFs accelerated computations for SI-GP-OPF by $70$ times while maintaining superior performance to the GP-OPF. The adopted Bayesian approach provided uncertainties that could flag unreliable learned setpoints.

This work sets the foundations for several exciting research directions. An interesting direction is to employ active learning to select the training data at locations with high uncertainty. This method can help reduce the training size while achieving high learning accuracy. Another practically relevant direction is to leverage spatiotemporal covariance between inverter dispatches and avoid learning the setpoints per inverter. SI-RF-GP-OPF models can also be used as digital twins or surrogates of the OPF in bilevel programming settings as they provide reasonable predictions for minimizers and their gradients alike.

\appendix
\subsection{Random Feature Approximation}\label{sec:AppA}
This appendix justifies the approximation in \eqref{eq:kernel_approx}. According to Bochner's theorem~\cite{Recht_RF}, every continuous, shift-invariant kernel $k(\btheta_i,\btheta_j)$ is the Fourier transform of a pdf $p(\bv)$ as
\begin{equation*}
k(\btheta_i,\btheta_j) = \int_{-\infty}^{+\infty}  e^{j\bv^\top (\btheta_i-\btheta_j)} p(\bv) \d\bv  = \mathbb{E}_{\bv} [ e^{j \bv^\top (\btheta_i-\btheta_j)}].
\end{equation*}
The Fourier transform of the Gaussian kernel in \eqref{eq:kernel} is known to be also a Gaussian yet of the inverse variance. Therefore, the pdf associated with the Gaussian kernel is $p(\bv)=\mcN(\bzero,\beta\bI_M)$. This is up to some scaling constants that can be absorbed into parameter $\alpha$. Leveraging this equivalence, one can draw $D$ samples $\{\bv_d\}_{d=1}^D$ from $\mcN(\bzero,\beta\bI_M)$, and compute a sample estimate of the earlier expectation as
\begin{equation}\label{eq:kernel_approx_complex}
\hat{k}(\btheta_i,\btheta_j):=\frac{1}{D}\sum_{d=1}^D e^{j\bv_d^\top\btheta_i}e^{-j\bv_d^\top\btheta_j}=\bzeta^\top(\btheta_j)\bzeta^*(\btheta_i) 
\end{equation}
where $^*$ denotes complex conjugation and $\bzeta(\btheta)$ is a $D$-length vector whose $d$-th entry is defined as $\zeta_d(\btheta):=e^{j\bv_d^\top\btheta}/\sqrt{D}$. Equation~\eqref{eq:kernel_approx_complex} provides an unbiased estimate of $k(\btheta_i,\btheta_j)$ and its variance decreases as $1/D^2$. To avoid working with complex-valued feature vectors, note that 
\begin{equation}\label{eq:c2r}
k(\btheta_i,\btheta_j)=\mathbb{E}_{\bv} [ e^{j \bv^\top (\btheta_i-\btheta_j)}]=\mathbb{E}_{\bv}[ \cos(\bv^\top (\btheta_i-\btheta_j))]
\end{equation}
This follows from Euler's identity and upon noting that $\mathbb{E}_{\bv}\left[ \sin\left(\bv^\top (\btheta_i-\btheta_j)\right)\right]=0$ because $\sin$ is an odd function and is applied to the zero-mean random variable $\bv^\top (\btheta_i-\btheta_j)$. The quantity $\cos(\bv^\top (\btheta_i-\btheta_j))$ can be expressed as the inner product between two $2D$-length vectors~\cite{MKL_RF}. To express this quantity as the inner product between $D$-length vectors as in \eqref{eq:kernel_approx_complex}, introduce an auxiliary random variable $\phi$ that is drawn uniformly from $[0,2\pi]$, so that \eqref{eq:c2r} can be expressed as~\cite{Recht_RF}
\begin{equation}\label{eq:c2r2}
k(\btheta_i,\btheta_j)=\mathbb{E}_{\bv,\phi}\left[ 2\cos\left(\bv^\top\btheta_i+\phi\right)\cos\left(\bv^\top\btheta_j+\phi\right)\right].  
\end{equation}
To see that \eqref{eq:c2r2} is equivalent to \eqref{eq:c2r}, use trigonometric identities and observe $\mathbb{E}_\phi[\cos(\bv^\top(\btheta_i+\btheta_j)+2\phi)]=0$ for all $\bv$. The approximation in \eqref{eq:kernel_approx} is a sample estimate of \eqref{eq:c2r2} obtained upon drawing $D$ samples of $(\bv_d,\phi_d)$.


\subsection{Building the Linear System of~\eqref{eq:SU}}\label{sec:AppB}
Matrices $\bS$ and $\bU$ in~\eqref{eq:SU} can be constructed from~\eqref{eq:td} as
\[\bS=\left[\begin{array}{cccc}
        \bA_e & \bzero & \bzero & \bzero \\
        \bS_{21} & \bA_e^\top & \bA_i^\top &\bS_{24}\\
        \diag(\bmu)\bA_i & \bzero & \bS_{33} & \bzero \\
        \bS_{41} & \bzero & \bzero & \bS_{44}
    \end{array}\right],~\bU {=}
        \left[\begin{array}{c}
        \bB_e\\
        \bzero\\
        \diag(\bmu)\bB_i\\
        \bzero    
        \end{array}\right]\] 
where the $m$-th column of $\bS_{24}$ is $\frac{\bA_m^\top\bA_m\bx}{\|\bA_m\bx\|}-\bb_m~\forall m$, and 
\begin{align*}
&\bS_{21}=\sum_{m=1}^{2N}\nu_m\left(\frac{\bA_m^\top\bA_m}{\|\bA_m\bx\|}-\frac{\bA_m^\top\bA_m\bx\bx^\top\bA_m^\top\bA_m}{\|\bA_m\bx\|^3}\right)\\
&\bS_{33}=\diag(\bA_i\bx-\bB_i\btheta-\bef_i)\\
&\bS_{41}=\diag(\bnu)\bS_{24}^\top,~~\bS_{44}=\diag(\{\|\bA_m\bx\|-\bb_m^\top\bx-f_m\}_{m=1}^{2N}).
\end{align*}

\subsection{Linearized OPF (LOPF)}\label{sec:LDF}
We cast an approximate OPF relying on a linearization of the power flow equations [cf.~\eqref{eq:LDF:v}]. This LOPF can be expressed as the quadratic program
\begin{subequations}\label{eq:LDF}
\begin{align}
\min~&~\bp^\top \bone + \bp^\top \bR\bp + \bq^\top \bR \bq\\
    \textrm{over}~&\{p_n^g, q_n^g\}_{n \in \mcN_g}, \{v_n\}_{n=1}^N, v_0\label{eq:LPF:x}\\
\textrm{s.to}~&~\bv = \bR \bp + \bX \bq+v_0 \bone\label{eq:LDF:v}\\
&~\eqref{eq:OPF:vmax},\eqref{eq:OPF:pmax}\\
&\left |p_n^g\cos\left(\tfrac{k\pi}{16}\right)+ q_n^g\sin\left(\tfrac{k\pi}{16}\right) \right| \leq \bar{s}_n^g,~k=1:16.\label{eq:LDF:poly}
\end{align}
\end{subequations}
where vectors $(\bp,\bq)$ collect the net power injections defined in \eqref{eq:pq}, and vector $\bv$ approximates nodal voltages as an affine function of injections with positive definite matrices $(\bR,\bX)$; see e.g., \cite{TJKT20}. The term $\bp^\top \bR\bp + \bq^\top \bR \bq$ is an approximation (second-order Taylor's series expansion in fact) of ohmic line losses~\cite[Prop.~1]{TJKT20}. Constraint~\eqref{eq:LDF:poly} approximates the quadratic constraint in~\eqref{eq:OPF:smax} using a 32-vertex polytope~\cite{Jabr18}. 

\bibliographystyle{IEEEtran}
\bibliography{myabrv,power}

\end{document}